\documentclass[12pt]{article}
\pdfoutput=1

\usepackage{jheppub}
\makeatletter
\def\@fpheader{\relax}
\makeatother

\usepackage{amsmath,amssymb}

\def\ba{\bar{a}}
\def\bb{\bar{b}}

\title{Probing the EVH limit of supersymmetric AdS black holes}

\author[a]{Kevin Goldstein}
\author[a]{\!\!, Vishnu Jejjala}
\author[a,b]{\!\!, Yang Lei}
\author[a]{\!\!, Sam van Leuven}
\author[c]{\!\!, Wei Li}
	
\affiliation[\,a]{Mandelstam Institute for Theoretical Physics, School of Physics, NITheP, and CoE-MaSS, \\ University of the Witwatersrand, 1 Jan Smuts Avenue, Johannesburg, South Africa}
\affiliation[\,b]{The Niels Bohr Institute, Copenhagen University,  Blegdamsvej 17, \\ DK-2100 Copenhagen \O , Denmark}
\affiliation[\,c]{Institute of Theoretical Physics, \\ Chinese Academy of Sciences, 100190 Beijing, P.R.~China}
		
\emailAdd{kevin.goldstein@wits.ac.za}
\emailAdd{vishnu@neo.phys.wits.ac.za}
\emailAdd{yang.lei@wits.ac.za}
\emailAdd{svleuven@xs4all.nl}
\emailAdd{weili@itp.ac.cn}

\abstract{
Extremal black holes in general dimensions are well known to contain an AdS$_2$ factor in their near-horizon geometries.
If the extremal limit is taken in conjunction with a specific vanishing horizon limit, the so-called  Extremal Vanishing Horizon (EVH) limit, the AdS$_2$ factor lifts to a locally AdS$_3$ factor with a pinching angular direction.
In this paper, we study the EVH limit of asymptotically AdS black holes which preserve some supersymmetry.
The primary example we consider is the 1/16$^{\rm th}$ BPS asymptotically AdS$_5$ black hole, whose EVH limit has an AdS$_3$ factor in its near-horizon geometry.
We also consider the near-EVH limit of this black hole, in which the near-horizon geometry instead contains an extremal BTZ factor.
We employ recent results on the large-$N$ limit of the superconformal index of the dual CFT$_4$ to understand the emergence of a CFT$_2$ in the IR of the CFT$_4$, which is the field theory dual to the emergence of the locally AdS$_3$ factor in the near-horizon geometry.
In particular, we show that the inverse Laplace transform of the superconformal index, yielding the black hole entropy, becomes equivalent to the derivation of a Cardy formula for the dual CFT$_2$.
Finally, we examine the EVH limit of supersymmetric black holes in other dimensions.
}

\date{}

\begin{document} 

\parskip=10pt

\maketitle 

\section{Introduction}

A remarkable realization of holography is the well-studied gauge/gravity correspondence \cite{Maldacena:1997re,Gubser:1998bc,Witten:1998qj}. 
String theory on AdS$_{d+1}\times X_{9-d}$, with $X_{9-d}$ compact, is dual to a $d$-dimensional conformal field theory living on the boundary of the anti-de Sitter space. 
The example of primary interest for us is the duality between type IIB string theory on AdS$_5 \times S^5$ and $\mathcal{N}=4$ super-Yang--Mills theory.
Avatars in other dimensions include: the duality between type IIB string theory on AdS$_3\times S^3\times X_4$ and the D$1$-D$5$ CFT$_2$ living on $X_4$ (for a review see \cite{David:2002wn}), 
the duality between type IIA string theory on AdS$_4\times \mathbb{CP}^3$ and the ABJM theory \cite{Aharony:2008ug},
and the duality between the $(0,2)$ superconformal field theory in six dimensions and M-theory on AdS$_7 \times S^4$ (see e.g.\ \cite{Mori:2014tca,Chester:2018dga,Heckman:2018jxk}).  

AdS/CFT provides a fundamental laboratory for the study of the physics of black holes.
More precisely, since the CFT provides a non-perturbative definition of string theory on AdS backgrounds, one can study questions about (asymptotically AdS) black holes in the well-defined setting of the quantum field theory.
The prototypical example is the BTZ black hole in AdS$_3$ \cite{Banados:1992wn,Banados:1992gq}, whose entropy can be understood in terms of the Cardy formula in the associated CFT$_2$ \cite{Strominger:1997eq,Mathur:2005zp,Hartman:2014oaa}.
Generic extremal black holes have an AdS$_2$ factor in their near-horizon geometry and have a vanishing temperature but finite horizon area \cite{Kunduri:2007vf}. 
Often, this is enhanced or lifted to an AdS$_3$ near-horizon, and the determination of the entropy utilizes known results about the BTZ geometry.

One lesson of AdS/CFT is that 
the entropy of a black hole can be calculated either from the point of view of the CFT corresponding to the AdS asymptopia (the UV CFT) or from a lower dimensional CFT corresponding to the near-horizon AdS factor (the IR CFT).
As entropy counts microscopic degrees of freedom, we would like to identify black hole microstates in both the UV CFT and the IR CFT and determine how one ensemble of states evolves smoothly to the other under the holographic renormalization group.
This is in general a challenging problem.
We specialize to settings in which algebraic features of the IR CFT are realized in terms of the UV CFT.

Recently, the entropy for certain supersymmetric asymptotically AdS$_4$ and AdS$_5$ black holes with finite horizon area has been accounted for from the dual CFT point of view \cite{Benini:2015eyy,Hosseini:2017mds,Benini:2018ywd,Cabo-Bizet:2018ehj,Choi:2018hmj,Kim:2019yrz,ArabiArdehali:2019tdm,Cabo-Bizet:2019osg,Honda:2019cio,Lezcano:2019pae,Larsen:2019oll,Nian:2019pxj,Azzurli:2017kxo,Bobev:2019zmz,Choi:2019zpz,Amariti:2019mgp,Lanir:2019abx}.
In this paper, we employ the results on $1/16^{\rm th}$ BPS AdS$_5$ black holes \cite{Gutowski:2004ez,Gutowski:2004yv,Chong:2005da,Chong:2005hr,Kunduri:2006ek} to study the so-called extremal vanishing horizon (EVH) black holes \cite{Balasubramanian:2007bs,Fareghbal:2008ar,Fareghbal:2008eh,SheikhJabbaria:2011gc,deBoer:2011zt,Johnstone:2013eg,Sadeghian:2015laa}.
The EVH black holes of interest to us are defined by first taking a limit of a generic (non-BPS) AdS$_5$ black hole in which the horizon area $A$ scales with a positive integer power of the temperature as $T\to 0$:
\begin{equation}\label{eq:EVHthermoST}
A \sim T^k, \qquad k>0.
\end{equation}
In fact, we restrict to the case $k=1$ in which, upon replacing $A$ with the entropy $S$, the thermodynamic relation \eqref{eq:EVHthermoST} is precisely what one expects for a CFT$_2$.
On the gravity side, it is then expected that a near-horizon (pinching) AdS$_3$ geometry emerges which decouples from the rest of the bulk directions.
This setup is what is known as the EVH/CFT$_2$ correspondence, where the CFT$_2$ represents the dual to the decoupled AdS$_3$ geometry.
We will also consider near-EVH black holes which have a (pinching) BTZ factor in their near-horizon geometry.
Such black holes arise when we in addition to the EVH limit take $G_3\to 0$, such that both the entropy and temperature are finite in the limit \cite{deBoer:2010ac,deBoer:2011zt,Johnstone:2013eg}.
These black holes correspond to excited states in the putative CFT$_2$. 

Our work aims to achieve a detailed understanding of the emergence of a CFT$_2$ in the IR from the CFT$_4$ in the UV from purely field theoretic considerations.\footnote{See also \cite{Berkooz:2014uwa,Berkooz:2012qh} for some results in this direction. Also, similar work has been done in \cite{Beem:2013sza,Basar:2015xda}.} 
In particular, we calculate the entropy for (near-)EVH black holes using the methods of \cite{Choi:2018hmj}.
These results apply to the $1/16^{\rm th}$ BPS AdS$_5$ black holes only, so we will restrict our attention to a (smooth) BPS limit of the EVH black holes.
One of our main findings is that the computation of the entropy in the four-dimensional CFT reduces to the derivation of the Cardy formula of a two-dimensional CFT with precisely the central charge and conformal dimension as expected from the EVH/CFT$_2$ proposal.
We take this as further evidence for the EVH/CFT correspondence derived from a purely field theoretical perspective.
We comment on how this derivation may help in understanding the field theory perspective in more detail.

Given the success of understanding the EVH limit of AdS$_5$ BPS black holes, it is interesting to know whether the method can also be used to study BPS black hole in other dimensions.
The entropy-temperature scaling definition \eqref{eq:EVHthermoST} of the EVH black holes requires knowledge about non-extremal AdS geometries at strong coupling.
Limitations of known AdS black hole solutions in general dimensions restricts our capacity to explore this further.
We find that among the set of known AdS$_{d+1}$ ($d>2$) black hole solutions, the AdS$_5$ BPS black hole  \cite{Johnstone:2013eg,Kunduri:2006ek,Chong:2005da} is the best example to explore the EVH-BPS limit. 

This paper is organized as follows.  
In Section \ref{sec:AdS5BH}, we review the most general black hole solutions in AdS$_5$ and subsequently restrict to supersymmetric solutions and study their (near-)EVH limits.
In Section \ref{sec:field}, we review the microscopic derivation of the supersymmetric AdS$_5$ black hole entropy.
We then use this calculation to derive the entropy of the (near-)EVH black holes and show how this computation is related to the derivation of the Cardy formula in two-dimensional CFT.
In Section \ref{sec:generaldimension}, we introduce a method to determine whether AdS$_{k+1}$ could appear in the near-horizon limit of EVH-BPS black holes in AdS$_{d+1}$ for $k<d$ for general $d$ and $k<d$. 
Finally, in Section \ref{sec:discussion} we discuss our results and conclude with directions for future research.

\section{Supersymmetric black hole solution in AdS$_5$} \label{sec:AdS5BH}

In this section, we will review the general AdS$_5$ black hole solutions.
In addition, we will take both the EVH and BPS limits of these solutions and derive their near-horizon geometries.
Finally, we review the constraint on the chemical potentials, which will be important to connect the geometry to its field theory dual.
The study of the EVH-BPS limit of black holes in general dimensions will be postponed to Section \ref{sec:generaldimension}.

\subsection{Black hole solutions in $5$d supergravity}\label{ssec:evh-limit-bh}

The aim of this section is to set notation.
As we write out the solutions explicitly, certain formulae are somewhat intricate.
We will first consider a supersymmetric AdS$_5$ black hole as the solution of five-dimensional gauged supergravity. 
The generic five-dimensional $U(1)^n$  gauged supergravity action  \cite{Kunduri:2006ek} is:\footnote{We set the coupling $g=1$.} 
\begin{equation}\label{eq:five-dimensionalaction}
S_{\text{5d}}= \frac{1}{16\pi G_5} \int \left(R_5 \star 1 -Q_{IJ} F^I \wedge \star F^J -Q_{IJ} dX^I \wedge \star dX^J -\frac{1}{6}C_{IJK} F^I \wedge F^J \wedge A^K +2 \mathcal{V} \star 1\right)\,,
\end{equation}
where the $n$ real $X^I$ obey the constraint 
\begin{equation}\label{eq:scalarconstraint}
\frac{1}{6}C_{IJK}X^IX^JX^K=1 \qquad \textrm{with} \quad I,J,K=1,\cdots, n\,,
\end{equation}
and $C_{IJK}$ is symmetric under permutation of $(IJK)$. In addition
\begin{equation}\label{eq:sugra2}
F^I\equiv dA^I\,, \qquad X_I\equiv \frac{1}{6}C_{IJK}X^JX^K \,,\qquad Q_{IJ}\equiv \frac{9}{2}X_IX_J-\frac{1}{2}C_{IJK}X^K\,.
\end{equation}

We will focus on $U(1)^3$ gauged supergravity, which contains $U(1)^3$ gauge fields and two independent real scalars.
Now $C_{IJK} =1$ when $(IJK)$ is a permutation of $(123)$ and zero otherwise.
In terms of $X^I$ with $I=1,2,3$, the constraint (\ref{eq:scalarconstraint}) and eq.\ (\ref{eq:sugra2}) become
\begin{eqnarray}
&& X^1 X^2 X^3 =1, \qquad X_I \equiv \frac{1}{3X^I}, \qquad Q_{IJ} \equiv \frac{\delta_{IJ}}{2(X^I)^2}.
\end{eqnarray}
Furthermore, the potential $\mathcal{V}$ is given by:
\begin{equation}
2 \mathcal{V} \star 1 = 4 \sqrt{-G}\sum_{i=1}^3\frac{1}{X^I}dV.
\end{equation}
The scalars $X^I$ can be written in terms of two independent scalars $\varphi_i$:  \begin{equation}
X^1 = e^{-\frac{1}{\sqrt{6}}\varphi_1-\frac{1}{\sqrt{2}}\varphi_2}, \quad X^2 = e^{-\frac{1}{\sqrt{6}}\varphi_1+\frac{1}{\sqrt{2}}\varphi_2}, \quad X^3 =e^{\frac{2}{\sqrt{6}}\varphi_1}.
\end{equation}
In terms of the $\varphi_i$, the bosonic part of Lagrangian of this $U(1)^3$ theory becomes \cite{Chong:2005da}: 
\begin{equation}\label{eq:five-dimensionalactionu13}
\frac{1}{\sqrt{|g|}}\mathcal{L} =R -\frac{1}{2}((\partial \varphi_1)^2+(\partial \varphi_2)^2 ) - \sum_{I=1}^3 \frac{1}{(2X^I)^2} (F^I)^2 +4 \sum_{I=1}^3 \frac{1}{X^I} +\frac{1}{6} \epsilon^{\mu\nu\rho\sigma\lambda} F^{1}_{\mu\nu} F_{\rho\sigma}^2 A_\lambda^3.
\end{equation}
The action associated to \eqref{eq:five-dimensionalactionu13} has a non-BPS black hole solution, described by four free parameters $(a,b,m,q)$ \cite{Chong:2005da}: 
\begin{equation}\label{eq:nonextremalBH}
\begin{aligned}
ds^2_5 &= H^{-\frac{4}{3}} \left[ -\frac{X}{\rho^2} (dt-a\sin^2 \theta \frac{d\phi}{\Xi_a}-b \cos^2 \theta \frac{d\psi}{\Xi_b})^2  + \frac{C}{\rho^2} \left( \frac{a b}{f_3} dt- \frac{b}{f_2} \sin^2 \theta \frac{d\phi}{\Xi_a} -\frac{a}{f_1} \cos^2 \theta \frac{d\psi}{\Xi_b}\right)^2 \right. \\ 
&+ \left. \frac{Z \sin^2\theta}{ \rho^2} \left(\frac{a}{f_3} dt-\frac{1}{f_2} \frac{d\phi}{\Xi_a}\right)^2+ \frac{W \cos^2\theta}{\rho^2} \left(\frac{b}{f_3}dt-\frac{1}{f_1} \frac{d\psi}{\Xi_b}\right)^2 \right] + H^{\frac{2}{3}} \left[ \frac{\rho^2}{X} dr^2 + \frac{\rho^2}{\Delta_\theta} d\theta^2 \right],\\ 
H&= \frac{\widetilde{\rho}^2}{\rho^2},\quad \rho^2 =r^2 +a^2\cos^2 \theta +b^2 \sin^2 \theta, \qquad \widetilde{\rho}^2 =\rho^2 +q ,\\
f_1&= a^2+r^2, \quad f_2 =b^2+r^2, \quad f_3 =(a^2+r^2 )(b^2+r^2) +qr^2 ,\\ 
\Delta_\theta &= 1-a^2 \cos^2\theta -b^2  \sin^2 \theta, \qquad X= \frac{(a^2+r^2)(b^2+r^2)}{r^2} -2m +(a^2+r^2+q) (b^2+r^2+q) ,\\
C &= f_1f_2(X+2m-\frac{q^2}{\rho^2}),  \qquad \Xi_a= 1-a^2, \quad \Xi_b =1-b^2,  \\ 
Z&= -b^2 C +\frac{f_2f_3}{r^2}\left[f_3 - r^2(a^2-b^2)(a^2+r^2+q)\cos^2 \theta \right] ,\\ 
W&= -a^2 C +\frac{f_1f_3}{r^2}\left[f_3 + r^2(a^2-b^2)(b^2+r^2+q)\sin^2 \theta \right] .
\end{aligned}
\end{equation}
The scalars are related to the warp factors in the metric:
\begin{equation}\label{eq:scalarsofnonsusybh}
X_1=X_2 =H^{-\frac{1}{3}}, \qquad X_3 =H^{\frac{2}{3}},
\end{equation}
and the gauge fields are  given by:
\begin{align}
\begin{split}
A^1&= A^2 = \frac{\sqrt{q^2+2mq}}{\widetilde{\rho}^2} \left( dt-a\sin^2\theta \frac{d\phi}{\Xi_a}-b\cos^2\theta \frac{d\psi}{\Xi_b} \right), \\
A^3&= \frac{q}{\rho^2}  \left( b\sin^2\theta \frac{d\phi}{\Xi_a} +a\cos^2\theta \frac{d\psi}{\Xi_b}\right).
\end{split}
\end{align}

The black hole solution is written in the asymptotic rotating frame. 
One can transform the solution to the asymptotic static frame by the coordinate transformation \cite{Chong:2005da}:
\begin{equation}
\phi^S =\phi +a t, \quad \psi^S=\psi +b t.
\end{equation}
In the asymptotic static frame, the chemical potentials and angular velocities are \cite{Chong:2005da}:
\begin{align}\label{eq:generalchemicalpotential}
\begin{split}
\Omega_a&= \frac{a(r_+^4+r_+^2b^2+r_+^2 q+b^2+r_+^2)}{(r_+^2 +a^2)(r_+^2+b^2)+q r_+^2}, \qquad \Omega_b=  \frac{b(r_+^4+r_+^2a^2+r_+^2 q+a^2+r_+^2)}{(r_+^2 +a^2)(r_+^2+b^2)+q r_+^2} , \\ 
\Phi_1&= \Phi_2 = \frac{\sqrt{q^2+2m q}r_+^2}{(r_+^2+a^2)(r_+^2+b^2)+qr_+^2}, \qquad \Phi_3 = \frac{q ab}{(r_+^2+a^2)(r_+^2+b^2)+qr_+^2}\,,
\end{split}
\end{align}
where the Newton constant $G_5$ is expressed in terms of the rank of the $SU(N)$ gauge group of the dual field theory by $\frac{\pi}{2G_5} =N^2$.
Moreover, $r_{+}$ is the outer horizon radius for which $X$ as defined in \eqref{eq:nonextremalBH} vanishes: $X(r_+)=0$.
In addition, the angular momenta, $U(1)$ charges, entropy, temperature, and mass of the  black hole \eqref{eq:nonextremalBH} are given by \cite{Johnstone:2013eg,Chong:2005da}: 
\begin{align}\label{eq:non-BPSdata}
\begin{split}
J_a &= \frac{N^2 a(2m +q\Xi_b)}{2 \Xi_b\Xi_a^2} , \qquad J_b =  \frac{N^2b(2m +q\Xi_a)}{2 \Xi_a\Xi_b^2}, \\ 
Q_1 &= Q_2 =\frac{N^2 \sqrt{q^2+2mq}}{2\Xi_a\Xi_b} , \qquad Q_3 = -\frac{N^2 a b q}{2\Xi_a\Xi_b}, \\ 
S&= \frac{N^2 \pi [(r_+^2+a^2)(r_+^2+b^2)+q r_+^2]}{\Xi_a \Xi_b r_+},\\ 
T_H &=  \frac{2r_+^6 +r_+^4 (1+a^2+b^2+2q)-a^2b^2}{2\pi r_+ [(r_+^2+a^2)(r_+^2+b^2)+q r_+^2]}, \\ 
E&= \frac{N^2}{4\Xi_a^2 \Xi_b^2}[2m(2\Xi_a+2\Xi_b-\Xi_a\Xi_b)+q (2\Xi_a^2+2\Xi_b^2+2\Xi_a \Xi_b-\Xi_a^2\Xi_b-\Xi_b^2 \Xi_a)] ,
\end{split}
\end{align}
where $J_a$ and $J_b$ measure the angular momenta around the $\phi$ and $\psi$ directions respectively,\footnote{
We adopt this notation because $J_{a}\equiv J_{\phi}$ vanishes when $a=0$, and $J_{b}\equiv J_{\psi}$ vanishes when $b=0$.} whereas $Q_I$ measure the electric charges of the black hole with respect to the $U(1)$ gauge fields $A^I$.

The black hole solution preserves a supersymmetry when its charges satisfy the following BPS condition \cite{Chong:2005da}:
\begin{equation}\label{eq:BPSequation}
E - J_a -J_b -Q_1-Q_2-Q_3 =0.
\end{equation}
In terms of the parameters $(a,b,m,q)$, this condition is equivalent to: \begin{equation}\label{eq:BPSq}
q= \frac{2m}{(a+b) (2+a+b)}.
\end{equation}
However, even when this constraint is satisfied, it turns out that the spacetime could contain closed timelike curves. 
To avoid those, and in addition naked singularities, one also has to constrain the parameter $m$ in terms of $a$ and $b$ \cite{Cvetic:2005zi}.
The result is:
\begin{equation}\label{eq:BPSm}
m= \frac{(a+b)^2(1+a)(1+b)(2+a+b)}{2(1+a+b)}.
\end{equation}
For these values, it turns out that the horizon radius is given by:
\begin{equation}\label{eq:r0ab}
r_+^2=r_0^2\equiv  \frac{ab}{1+a+b}.
\end{equation} 
Plugging in these values, the function $X$ simplifies to: 
\begin{equation}\label{eq:Xr}
X(r) = \frac{(r^2-r_0^2)^2 ( r^2 +(1+a+b)^2)}{r^2},
\end{equation}
showing that the solution has become extremal. 
Namely, a causally well-behaved BPS black hole is also extremal. 
At the BPS-extremal point, the charge $q$ simplifies to:
\begin{equation}\label{eq:qBPS}
q= \frac{(a+b)(1+a)(1+b)}{1+a+b}.
\end{equation}
Therefore, this BPS black hole only depends on the parameters $a$ and $b$.  

The supersymmetric solution  we study here looks different from the most general supersymmetric black holes found in  \cite{Kunduri:2006ek}, which we review in Appendix \ref{sec:generalEVHBH}. 
The supersymmetric solution in \cite{Kunduri:2006ek} chooses a $t$ coordinate adapted to the null Killing field $\partial_t$ on the horizon. 
Thus, the solution \eqref{eq:susyBHansatz} is corotating with horizon.\footnote{We thank James Lucietti for clarifying this point for us.} 
To embed the supersymmetric limit of solution \eqref{eq:nonextremalBH} into the most general class \eqref{eq:susyBHansatz}, we need to transform the metric into the corotating frame via: 
\begin{equation}\label{eq:corotatingco}
\widetilde{\phi} = \phi +(a-1)t, \quad \widetilde{\psi}= \psi+(b-1) t.
\end{equation}

Finally, the black hole solution \eqref{eq:nonextremalBH} can be embedded into ten-dimensional type IIB supergravity \cite{Cvetic:1999xp} and also into eleven dimensions supergravity \cite{Colgain:2014pha}. 
The ansatz for the ten-dimensional supergravity solution is 
\begin{equation}\label{eq:ten-dimensionalembedding}
ds_{10}^2 =\sqrt{\widetilde{\Delta}} ds_5^2 +\frac{1}{\sqrt{\widetilde{\Delta}}} \sum_{i=1}^3 \frac{1}{X_i} (d\mu_i^2+\mu^2_i(d\psi_i+A^i)^2).
\end{equation}
The extra five-dimensional manifold can be considered as a deformation of $S^5$, which is parametrized by:
\begin{equation}
d\Omega_{S^5}^2 = \sum_i \left( d\mu_i^2+\mu^2_id\psi_i^2 \right).
\end{equation}
Here, we think of $S^5$ as a three-torus $T^3$ fibration over the two-sphere $S^2$, where the $\mu_i$ parametrize the $S^2$: 
\begin{equation}\label{eq:mu-alpha}
\mu_1=\sin \alpha \cos\beta, \quad \mu_2 = \sin\alpha \sin \beta, \quad \mu_3 = \cos \alpha .
\end{equation}
On the other hand, the $\psi_i$ parametrize $T^3$. Finally, $\widetilde{\Delta}$ only depends on the scalar fields $X_i$ through $\widetilde{\Delta} = \sum_i \mu_i^2 X_i$.

\subsection{(Near-)EVH limit of the BPS black hole}\label{ssec:near-EVH-bhs}

The BPS black hole we constructed above generically has a finite horizon area. 
In particular, its near-horizon geometry contains an AdS$_2$ throat \cite{Hosseini:2017mds}. 
In order for an AdS$_3$ to arise in the near-horizon geometry, the horizon size of the black hole has to vanish.
The AdS$_3$ geometry reflects the vanishing horizon area in that its angular direction is pinching.
In general, this will result into an infinitely gapped system, since the energy gap above the ground state scales like $1/R$, with $R$ the (conformal) boundary circle radius.
However, as studied in \cite{deBoer:2010ac,deBoer:2011zt}, one may keep non-trivial dynamics by combining the vanishing entropy limit with a $G_3\to 0$ (\textit{i.e.}, $c\to \infty$) limit.
This will be the type of double scaling limit we consider for the near-EVH black holes.

Let us first revisit the strict EVH black hole obtained in \cite{Johnstone:2013eg}. 
The EVH condition can be imposed independently of the supersymmetry condition. 
For the solution \eqref{eq:nonextremalBH} with a ten-dimensional embedding \eqref{eq:ten-dimensionalembedding}, the EVH condition is 
\begin{equation}
X(r_+=b=0)=0.
\end{equation}
This gives: 
\begin{equation}
m=\frac{q^2+a^2(1+q)}{2}.
\end{equation}
The near-horizon limit is approached by expanding \eqref{eq:nonextremalBH} in terms  of $r=\epsilon \rho$.
After some work and suitable coordinate redefinitions, \cite{Johnstone:2013eg} obtains the following geometry:
\begin{align}\label{eq:EVHnearhorizon}
\begin{split}
ds^2 &= h_1h_2 \left[-\frac{x^2}{\ell^2_3} d\tau^2 + \frac{\ell^2_3}{x^2}dx^2+x^2d\widetilde{\chi}^2\right] \\ 
&+ \frac{(a^2+q)h_1h_2}{\Delta_\theta} d\theta^2+ \frac{\cos^2\alpha \cos^2\theta }{K^2 h_1h_2}d\xi^2 + \frac{a^2+q}{\Xi_a^2}\frac{h_2}{h_1^3}\Delta_\theta \sin^2\theta d\widetilde{\phi}^2  \\
&+ \frac{h_2}{h_1}d\alpha^2+\frac{h_1}{h_2}\sin^2\alpha\, d\beta^2 +\frac{h_1}{h_2}\left[\mu_1^2(d\widetilde{\psi}_1-Ad\widetilde{\phi})^2+\mu_1^2(d\widetilde{\psi}_2-Ad\widetilde{\phi})^2\right].
\end{split}
\end{align}
Apart from the definitions of the various functions that we will collect below, let us start by summarizing the most important aspects of this geometry.
First of all, the metric describes a warped product of a locally AdS$_3$ geometry (first line \eqref{eq:EVHnearhorizon}) and some seven-dimensional manifold (second and third line in \eqref{eq:EVHnearhorizon}).
The time and radial coordinates $\tau$ and $x$ are directly related to the time and radial coordinates of the original AdS$_5$ black hole geometry. 
In addition, the new angular coordinates $\widetilde{\chi}$ and $\xi$ are related to the original angles $\psi$ and $\phi$ of the AdS$_5$ geometry via:
\begin{equation}\label{eq:ang-coord-transf}
\widetilde{\chi}=\epsilon\psi,\qquad \xi=\frac{1}{\sqrt{2}}(\psi_3+\psi).
\end{equation}
This implies in particular that the periodicity of $\widetilde{\chi}$ is $2\pi \epsilon$, showing the local AdS$_3$ geometry is the pinching orbifold of \cite{deBoer:2010ac}.
Moreover, the coordinate transformation shows that the AdS$_3$ geometry is embedded completely inside the AdS$_5$ geometry, as opposed to its angular part originating from an $S^5$ direction as in \textit{e.g.}, \cite{Balasubramanian:2007bs}.
Finally, note that the generator of rotations around the angular direction in AdS$_3$ will be proportional to $\partial_{\widetilde{\chi}}= \epsilon^{-1}(\partial_\psi+\partial_{\psi_3})$.\footnote{This will explain a relation we discuss below in the context of the near-EVH black hole, which relates the $L_0$ eigenvalue of the CFT$_2$ to the near-EVH black hole charges $L_0\sim J_b+Q_3$ (since in the BPS case $\bar{L}_0=0$).}

Let us now briefly collect the definitions of the various functions appearing in the metric.
Some were already defined in \eqref{eq:nonextremalBH} and \eqref{eq:mu-alpha}.
For the rest, $h_{1,2}$ are given by \cite{Johnstone:2013eg}:
\begin{equation}
h_1^2 =\frac{a^2\cos^2\theta +q}{a^2+q}, \qquad h_2^2 = \frac{a^2\cos^2\theta +q\mu_3^2}{a^2+q}.
\end{equation}
Furthermore, we have:
\begin{equation}
\Delta_\theta=1-a^2\cos^2\theta,\qquad A=\frac{a\sqrt{q\mathbf{Y}_s}}{\ell_3\Xi_a\sqrt{\mathbf{V}}}\frac{\sin^2\theta}{h_1^2}.
\end{equation}
Finally, we have:
\begin{equation}
K=\sqrt{\frac{a^2+q}{a^2+q^2}},\qquad \ell_3^2=\frac{a^2+q}{\mathbf{V}}.
\end{equation}
At the BPS point, when we are imposing conditions \eqref{eq:BPSq} and \eqref{eq:BPSm}, $\mathbf{V}$, $q$ and $\mathbf{Y}_s$ take the values: 
\begin{equation}
\mathbf{V}= (1+a)^2, \quad q=a, \quad \mathbf{Y}_s =1+a.
\end{equation}

Finally, let us write down the charges of this specific black hole.
First of all, note that $b=0$ implies that $J_b=Q_3=0$. 
The other charges are finite and given by: 
\begin{equation}\label{eq:JaQEVH}
J_a= \frac{N^2 a^2}{2(1-a)^2}, \quad Q_1= Q_2 = \frac{a}{2(1-a)}N^2,
\end{equation}
and satisfy:
\begin{equation}\label{eq:identifiyJQsqure}
\frac{N^2}{2}J_a = Q_1^2.
\end{equation}
We will come back to this relation in Section \ref{sec:field}.

To obtain the near-horizon metric for the near-EVH black hole, instead of $b=0$ one takes $b=\lambda \epsilon^2$.
The analysis that leads to the near-horizon geometry is otherwise similar to the strict EVH case, so we will not repeat it here and refer the interested reader to \cite{Johnstone:2013eg}.
Suffice to say that the near-horizon geometry describes a similar warped product as in \eqref{eq:EVHnearhorizon} but now instead of an AdS$_3$ geometry there is a pinching extremal BTZ in the BPS case:
\begin{equation}\label{eq:pinchingBTZ}
ds_{3}^2 = - \frac{\left(x^2-x_0^2\right)^2}{\ell_3^2 x^2} d\tau^2 + \frac{\ell_3^2x^2 dx^2}{(x^2-x_0^2)^2}  +x^2  \left(d\widetilde{\chi}-\frac{x_0^2}{\ell_3^2 x^2}d\tau\right)^2,
\end{equation}
where:
\begin{equation}
\ell_3^2 = \frac{a}{1+a}, \qquad x_0^2 = \frac{\lambda}{2 }.
\end{equation}
Moreover, the periodicity of $\widetilde{\chi}$ is again $2\pi \epsilon$, reflecting the fact that the BTZ is pinching.

By the relation between the parent central charge of the $\mathcal{N}=4$ theory to the ten-dimensional Newton constant and compactifying the warped geometry to the BTZ in three dimensions, we obtain \cite{Johnstone:2013eg}:
\begin{equation}
\frac{1}{G_3}=2\sqrt{2}N^2\frac{a}{1-a}\sqrt{\frac{a}{1+a}},
\end{equation}
with $N$ the rank of the gauge group.
This allows us to compute the entropy of the BTZ black hole:
\begin{equation}\label{eq:sbtz}
S_{\rm BTZ}=\frac{2\pi\epsilon x_0}{4G_3}= \frac{\pi a}{1-a} \sqrt{\frac{\lambda a}{1+a}} N^2 \epsilon .
\end{equation}
As mentioned at the beginning of this section, we see from this formula that to allow for a finite entropy we need to take the double scaling limit $\epsilon\to 0$ and $N\to\infty $ keeping $N^2\epsilon$ fixed.
This then defines the near-EVH limit.
We can also read off the mass of the BTZ black hole.
It is given by:
\begin{equation}\label{eq:mbtz}
M_{\rm BTZ}\ell_3= L_0 -\frac{c}{24}=\frac{a\lambda}{2\sqrt{2}(1-a)}N^2\epsilon = \frac{1}{\sqrt{2}\epsilon} (J_b +Q_3).
\end{equation}
In the first equality, we remind the reader how the (extremal) BTZ mass is related to the conformal dimension in the dual CFT$_2$, whereas the last equality can be checked by plugging in $b=\lambda\epsilon^2$ into the black hole charges \eqref{eq:non-BPSdata}.
Note that we anticipated this relation below \eqref{eq:ang-coord-transf}.
We will give a precise CFT$_4$ interpretation of both \eqref{eq:sbtz} and \eqref{eq:mbtz} in Section \ref{sec:field}.

Let us end this section by noting that our near-EVH black hole geometry \eqref{eq:pinchingBTZ} has two free parameters $(a,\lambda)$. 
The most general supersymmetric AdS$_5$ black hole \cite{Kunduri:2006ek} has four free parameters, which still has a vanishing entropy if $Q_3=J_b=0$ \cite{Choi:2018hmj}. 
This implies the EVH black hole also exists in these solutions. 
We will show in Appendix \ref{sec:generalEVHBH} that the near-horizon geometry of these most general supersymmetric AdS$_5$ black holes still contain AdS$_3$ factors in the EVH limit.

\subsection{Chemical potentials} \label{ssec:complex-chem-pots}

In this section, we will define the various potentials for the supersymmetric black holes by taking the BPS limit of the potentials for the generic non-BPS black hole \eqref{eq:generalchemicalpotential}.
The BPS limit can be taken through a zero temperature limit accompanied by specific limits for the various potentials as discussed in \cite{Silva:2006xv} (see also \cite{Cabo-Bizet:2018ehj} for a more recent discussion in our context). 
The fact that one has to take limits of the chemical potentials as well can be illustrated by a simple example \cite{Silva:2006xv}.
Consider the partition function
\begin{equation}
Z = \sum_{E,J} e^{-\beta E-\beta \Omega J} = \sum e^{-\beta (E-J)+\beta(1-\Omega)J},
\end{equation}
where $\beta$ is the inverse temperature and $\Omega$ is the chemical potential associated to the charge $J$. 
In the zero temperature limit $\beta \to \infty$, we see that the partition function localizes on the states satisfying a BPS equation $E=J$.
However, for the partition function to be well defined, one should also make sure that $\Delta \equiv\beta(1-\Omega)$ is finite in the limit $\beta\to\infty$. 
It is $\Delta$ that then defines the BPS limit of the chemical potential. 

To find the BPS values of the chemical potentials, we follow \cite{Cabo-Bizet:2018ehj}. 
We can start from the non-BPS black hole \eqref{eq:nonextremalBH}. 
The parameter $m$ can be expressed in terms of the horizon size $r_+$ as:
\begin{equation}
m= \frac{(r_+^2 +a^2)(r_+^2 +b^2)}{2r_+^2} + \frac{(r_+^2 +a^2+q)(r_+^2 +b^2+q)}{2}.
\end{equation} 
Imposing the supersymmetry condition \eqref{eq:BPSq} without the causality condition \eqref{eq:BPSm}, we can solve for $q$ as:\footnote{We have chosen one branch of the solution. The other can be obtained by $i\rightarrow -i$.}
\begin{equation}\label{eq:q-complex}
q =\frac{(a-i r_{+})(b-ir_{+})(1-i r_{+})}{-ir_{+}}= a+b+ ab -r_+^2 -\frac{i(1+a+b)}{r_+} (r_+^2 -r_0^2).
\end{equation} 
The parameter $q$ is only real when $r_+ =r_0$ with $r_0$ determined by \eqref{eq:r0ab}, which reproduces the BPS-extremal value (\ref{eq:qBPS}).
For all the other values, the BPS black hole has a naked singularity and is causally pathological \cite{Chong:2005da}. 
However, to take the BPS limit it turns out to be convenient to momentarily ignore the constraint $r_+ =r_0$ and take $q$ as in \eqref{eq:q-complex}.
We can then write the black hole chemical potentials $\Phi_I$ and angular velocities $\Omega_i$ in \eqref{eq:generalchemicalpotential} as follows:
\begin{align}\label{eq:complex-chem-pots}
\begin{split}
\Phi_1 &= \Phi_2 = \frac{(1+a+b )r_+(r_++i)}{i(1+a+b)r_++ab}, \qquad \Phi_3 = \frac{ab (i+r_+)}{r_+(ab+i(1+a+b)r_+)} ,\\
\Omega_a &= \frac{a(1-ir_+)[(1+a+b)r_+-i b]}{(a-ir_+)[(1+a+b)r_+-i a b]}, \qquad \Omega_b = \frac{b(1-ir_+)[(1+a+b)r_+-i a]}{(b-ir_+)[(1+a+b)r_+-i a b]}.
\end{split}
\end{align}
One verifies that these potentials obey the equation:
\begin{equation}
\beta(\Omega_a +\Omega_b -\Phi_1 -\Phi_2 -\Phi_3+1) =2\pi i,
\end{equation}
where $\beta$ corresponds to the inverse Hawking temperature of the non-BPS black hole given in \eqref{eq:non-BPSdata}.
We now want to consider the $\beta\to \infty$ limit, obtained by taking $r_+\to r_0$ with $r_0$ defined in \eqref{eq:r0ab}.
The chemical potentials in this limit simplify to:
\begin{equation}
\Phi_1 = \Phi_2 = \Phi_3 =\Omega_a=\Omega_b=1.
\end{equation}
As in the simple example discussed above, this motivates us to define finite BPS potentials as:
\begin{equation}\label{eqomegadelta}
\omega_i = \lim_{r_+\to r_0}\beta(1 -\Omega_i), \quad \Delta_I =\lim_{r_+\to r_0}\beta(1 -\Phi_I),
\end{equation}
with $i=a,b$ and $I=1,2,3$.
Explicitly, $\omega_i$ and $\Delta_I$ are given by:
\begin{align}\label{eq:bps-potentials}
\begin{split}
\Delta_1 &= \Delta_2 = -\frac{\pi r_0(a+b)[(1+a+b)r_0+iab]}{iab(1+a+b-ir_0)}, \\ \Delta_3 &= \frac{\pi (a+b)[(1+a+b)r_0+iab]}{ab(1+a+b-ir_0)},\\
\omega_a &= \frac{\pi(1-a)(ib+r_0)}{ir_0(1+a+b-ir_0)}, \qquad \omega_b =  \frac{\pi(1-b)(ia+r_0)}{ir_0(1+a+b-ir_0)}.
\end{split}
\end{align}
The BPS potentials then satisfy the constraint:
\begin{equation}\label{eq:contraintdeltaomega}
\Delta_1+  \Delta_2+ \Delta_3- \omega_a-\omega_b = \beta  -\beta(\Phi_1+\Phi_2+\Phi_3-\Omega_a-\Omega_b) =2\pi i.
\end{equation}
Therefore, even though the original chemical potentials \eqref{eq:complex-chem-pots} become real in the limit $r_+=r_0$, the properly defined BPS potentials satisfy a complex constraint!
This observation underlies all recent derivations of the AdS$_5$ BPS black hole entropy from the supersymmetric partition function of the $\mathcal{N}=4$ SYM theory \cite{Cabo-Bizet:2018ehj,Choi:2018hmj,Benini:2018ywd}.

In Section \ref{sec:field}, we will derive the values of the chemical potentials in the (near-) EVH limit from the dual CFT.
Therefore, we will now take the EVH limit of these potentials in order to compare to the field theory analysis later.
It turns out that in this comparison there is a slight subtlety which is related to the fact that in the strict EVH limit $b\to 0$, the EVH limit and BPS limit of the chemical potentials do not commute.
To see this, let us first compute the EVH limit of the BPS potentials in \eqref{eqomegadelta}:
\begin{equation}\label{eq:valueofcem}
\lim_{b\to 0}\, \Delta_1 =\lim_{b\to 0}\, \Delta_2 = \frac{\pi i a}{1+a} \,, \qquad   \lim_{b\to 0} \, \omega_a = -\frac{\pi i (1-a)}{1+a}.
\end{equation}
This shows that in the EVH limit $\Delta_{1,2}$ and $\omega_a$ satisfy the following condition:
\begin{equation}\label{eq:constraintEVHlimit}
\Delta_1+\Delta_2 -\omega_a = \pi i.
\end{equation}
In addition:
\begin{equation}\label{eq:BPSomegab}
\lim_{b\to 0}\,\omega_b= \frac{\pi}{\sqrt{b}} \sqrt{\frac{a}{1+a}}+\mathcal{O}(b^0).
\end{equation}

On the other hand, if we first take the EVH limit, such that in particular $J_b=Q_3=0$, the BPS condition \eqref{eq:BPSequation} reads:
\begin{equation}
E= Q_1+Q_2+J_a
\end{equation}
This condition may be viewed as the consequence of imposing two independent BPS conditions:
\begin{align}
\begin{split}
E &= Q_1+Q_2 +Q_3+J_a +J_b, \\
E &= Q_1+Q_2 -Q_3 +J_a -J_b.
\end{split}
\end{align}
For the black hole solutions we have studied, the first condition results in the constraint \eqref{eq:contraintdeltaomega} for generic $r_+$, as we have seen above.
On the other hand, a similar analysis shows that black hole solutions obeying the second condition have chemical potentials that obey another constraint:
\begin{equation}
\Delta_1+  \Delta_2- \Delta_3- \omega_a+\omega_b=2\pi i.
\end{equation}
These two constraints combine to give:
\begin{equation}\label{eq:EVHblackconstraint}
\Delta_1+\Delta_2 -\omega_a =2 \pi i,
\end{equation}
which is in clear contradiction to \eqref{eq:constraintEVHlimit}.
The explicit potentials we derive for $b\to 0$ in this case are given by:
\begin{equation}\label{eq:valueofcem2}
\lim_{b\to 0}\, \Delta_1 =\lim_{b\to 0}\, \Delta_2 = \frac{2\pi i a}{1+a} \,, \qquad   \lim_{b\to 0} \, \omega_a = -\frac{2\pi i (1-a)}{1+a}.
\end{equation}
As we will see in the next section, from the field theory point of view the latter potentials emerge in the strict EVH limit.
However, in the near-EVH limit we do find the values \eqref{eq:valueofcem} and \eqref{eq:BPSomegab}.

We will call the black holes satisfying the constraint \eqref{eq:constraintEVHlimit} the BPS black hole in the EVH limit. 
On the other hand, the black holes satisfying the constraint \eqref{eq:EVHblackconstraint} is called the EVH-BPS black hole. 
The constraint \eqref{eq:EVHblackconstraint} is the result of enhancement of supersymmetry in the strict EVH limit, as we will see in the following section.

\section{$\mathcal{N}=4$ SYM and the superconformal index}\label{sec:field}

To study the microscopic origin of the EVH/CFT correspondence, we now turn to the CFT dual description of the AdS$_5$ black holes discussed in the previous section.
This dual description is provided by the four-dimensional $\mathcal{N}=4$ $SU(N)$ super-Yang--Mills theory.
One of the most basic entries in the dictionary is the relation between the rank of the gauge group and the AdS radius and Newton's constant in five dimensions:
\begin{equation}
N^2=\frac{\pi }{2G_5g^3},
\end{equation}
where we put $g=L^{-1}_{AdS}=1$ in the following. 
To arrive at a microscopic derivation of the entropy of the (near-)EVH black holes, we will follow the recent work on a microscopic accounting of the entropy of general AdS$_5$ black holes \cite{Choi:2018hmj}.\footnote{See also \cite{Cabo-Bizet:2018ehj,Benini:2018ywd} for similar results which were obtained through different methods.}

The partition function studied in \cite{Choi:2018hmj} is defined by:
\begin{equation}\label{eq:gen-part-funct-choi}
Z(\beta,\Delta_I,\omega_i)=\text{Tr}_{\mathcal{H}}\left[ e^{-\sum^3_{I=1}\Delta_I Q_I}e^{-\omega_a J_a}e^{-\omega_b J_b}e^{-\beta E_{\rm susy}} \right].
\end{equation}
In this expression, $\mathcal{H}$ is the Hilbert space of the theory quantized on $S^3$ (the conformal boundary of AdS$_5$), which via the operator-state correspondence is constructed by applying gauge invariant local operators on the vacuum.
Furthermore, the $Q_I$ charges correspond to the Cartan generators of the $SO(6)$ R-symmetry and the $J_i$ charges correspond to the Cartan generators of the $SO(4)$ rotational symmetry of $S^3$.
The operator $E_{\rm susy}$ is taken to be:
\begin{equation}
E_{\rm susy}\equiv\lbrace \mathcal{Q},\mathcal{S}\rbrace=E-\left(\sum^3_{I=1}Q_I\right)-J_a-J_b\,,
\end{equation}
where $\mathcal{Q}$ is a supercharge with charge $+\frac{1}{2}$ under the $Q_I$ and charge $-\frac{1}{2}$ under the $J_i$, and $E$ is the generator of time translations of the theory on $S^3$.
Finally, the chemical potentials $\Delta_I$ and $\omega_i$ can in general be complex, as we will see below, but are periodic under translations by $4\pi i$.

As it stands, the partition function in \eqref{eq:gen-part-funct-choi} receives contributions from all states in the theory.
This is because $\mathcal{Q}$ does not anti-commute with the operator in the trace.
Indeed, we have:
\begin{equation}
e^{-\Delta_I Q_I-\omega_i J_i}\mathcal{Q}=e^{-\frac{\left(\sum^3_{I=1}\Delta_I\right)-\omega_a-\omega_b}{2}}\mathcal{Q}e^{-\Delta_I Q_I-\omega_i J_i}.
\end{equation}
To resolve this issue, it is proposed in \cite{Choi:2018hmj} (see also \cite{Cabo-Bizet:2018ehj}) to restrict the theory to the hypersurface:
\begin{equation}\label{eq:hypersurface-chem-pots}
\left(\sum^3_{I=1}\Delta_I\right)-\omega_a-\omega_b=2\pi i \mod 4\pi i. 
\end{equation}
Note that this is also the constraint obeyed by black holes as reviewed in Section \ref{ssec:complex-chem-pots}.
On this hypersurface, we see that the supercharge $\mathcal{Q}$ anticommutes with the full operator in the trace and therefore the partition function localizes on the space of 1/16$^{\rm th}$ BPS states:
\begin{equation}
\mathcal{H}_{\mathcal{Q}}=\lbrace |\psi\rangle\in \mathcal{H}|\quad \mathcal{Q} |\psi\rangle=0 \rbrace.
\end{equation}
This is achieved without the need for an explicit insertion of $(-1)^F$.
Since these BPS states are annihilated by $\lbrace\mathcal{Q},\mathcal{S}\rbrace$ as well, their quantum numbers saturate the unitarity bound:
\begin{equation}\label{eq:1/16bps-cond}
E\geq Q_1+Q_2+Q_3+J_a+J_b.
\end{equation}

In \cite{Choi:2018hmj}, the index \eqref{eq:gen-part-funct-choi} is explicitly computed in the weakly coupled $\mathcal{N}=4$ theory, even though the final result holds for arbitrary coupling due to the fact that only BPS states contribute.
As is standard, one deals with the gauge invariance constraint on states by including holonomies for the gauge field along the temporal circle in the trace \eqref{eq:gen-part-funct-choi}.
The projection on the gauge invariant states can then be performed by integration over the gauge group with the Haar measure:
\begin{equation}
\int [dU] = \frac{1}{N!(2\pi)^{N-1}} \int\prod_{i=1}^{N-1} d\alpha_i \prod_{i<j} 4\sin^2 \frac{1}{2} (\alpha_i-\alpha_j).
\end{equation}
The resulting expression for the index is given by \cite{Dolan:2008qi,Choi:2018hmj}:
\begin{equation}\label{eq:zerotemperaturepartitionlimi}
Z =\frac{1}{N!} \int \prod_{i=1}^{N} \frac{d\alpha_i}{2\pi} \prod_{i<j} \left(2\sin \frac{\alpha_{ij}}{2}\right)^2 \exp\left[ \sum_{n=1}^\infty \frac{1}{n} \left(1-\frac{\prod_{I=1}^{3} 2\sinh \frac{n \Delta_I}{2}}{2\sinh \frac{n\omega_a}{2}2\sinh \frac{n\omega_b}{2}} \right) \sum_{i,j=1}^N e^{in \alpha_{ij}}\right].
\end{equation}
Here, the $\alpha_i$ correspond to the weights of $SU(N)$ and $\alpha_{ij}\equiv \alpha_i-\alpha_j$ to the roots, which reflect the gauge charges of the various adjoint valued fields.

At large-$N$, this index can be computed through a saddle point approximation. 
The early result of \cite{Kinney:2005ej} showed that the leading term is only of order $\mathcal{O}(1)$ at large-$N$. 
The failure to reproduce an $\mathcal{O}(N^2)$ scaling, as expected from the existence of large AdS$_5$ BPS black holes, is believed to be due to the large cancellation between bosonic and fermionic degrees of freedom.

The crucial ingredient of the new analysis in \cite{Choi:2018hmj} is to allow for complex chemical potentials (as also emphasized in \cite{Cabo-Bizet:2018ehj,Benini:2018ywd}). 
This is in fact required since the black hole chemical potentials satisfy a complex constraint \eqref{eq:hypersurface-chem-pots}. 
The complex phase could potentially obstruct the large cancellation between bosonic and fermionic degrees of freedom. 
This turns out to be the case.
The leading order contribution to the partition function \eqref{eq:zerotemperaturepartitionlimi} at large-$N$ is computed through a saddle point approximation in the ``generalized Cardy'' limit $\omega_i\ll 1$ and is given by the simple $\mathcal{O}(N^2)$ expression:
\begin{equation}\label{eq:large-n-gen-partn-function}
\log Z= \frac{N^2}{2} \frac{\Delta_1\Delta_2\Delta_3}{\omega_a\omega_b}.
\end{equation}
To arrive at this expression, it is assumed that the matrix integral in \eqref{eq:zerotemperaturepartitionlimi} is dominated by the ``maximally deconfining'' saddle $\alpha_1=\alpha_2=\ldots=\alpha_N$. 
Note that the expression is formally identical to the supersymmetric Casimir energy \cite{Hosseini:2017mds,Assel:2014paa,Assel:2015nca}, but with chemical potential subject to the constraint (\ref{eq:hypersurface-chem-pots}).
To compute the degeneracy of states for fixed charges, one should now compute the inverse Laplace transform: 
\begin{equation}\label{eq:inted}
d = \int d\Delta_1d\Delta_2 d\omega_a d\omega_b\, Z(\Delta_I,\omega_i)\, e^{\omega_a(J_a+Q_3)}e^{\omega_b(J_b+Q_3)} e^{\Delta_1(Q_1-Q_3)} e^{\Delta_2(Q_2-Q_3)}(-1)^F,
\end{equation}
where $(-1)^F = e^{2\pi i Q_3}$. 
At large-$N$ and large charges (\textit{i.e.}, the charges scale with $N^2$), the inverse Laplace transform becomes a Legendre transform of $\log Z$. 
As shown in \cite{Hosseini:2017mds} (see also \cite{Cabo-Bizet:2018ehj,Choi:2018hmj}), this reproduces the expected BPS AdS$_5$ black hole entropy: 
\begin{equation}\label{eq:blackholeentropy0}
S = 2\pi \sqrt{Q_1 Q_2 +Q_2 Q_3 +Q_3 Q_1 -\frac{N^2}{2}(J_a+J_b)}.
\end{equation}

\subsection{Entropy of (near-)EVH black holes}\label{ssec:entropy-near-EVH}

We now turn to our black holes of interest, as described in Section \ref{ssec:near-EVH-bhs}.
This will not be a trivial specialization of the analysis reviewed above, since for the near-EVH black hole the charges $J_b$ and $Q_3$ are not of order $N^2$ and therefore it is not directly clear whether one can perform a saddle point approximation with respect to the associated chemical potentials.

\paragraph{The EVH limit:}

Since (near-)EVH black holes have $Q_1=Q_2\equiv Q$, the expression \eqref{eq:inted} becomes: 
\begin{equation}\label{eq:inted12}
\begin{aligned}
d &=\int d\Delta_1d\Delta_2 d\omega_a d\omega_b \exp\Bigl(\frac{N^2}{2}\frac{\Delta_1\Delta_2 \Delta_3}{\omega_a\omega_b} \\
&\qquad \qquad+\omega_a(J_a+Q_3) +\omega_b(J_b+Q_3) +(\Delta_1+\Delta_2)(Q-Q_3) +2\pi i Q_3\Bigr)\,.
\end{aligned}
\end{equation}
The saddle point equations for $\Delta_1$ and $\Delta_2$ are identical in this case, implying $\Delta_1=\Delta_2\equiv \Delta$ on the saddle.
In the following, we will use this fact to write \eqref{eq:inted12} as:
\begin{equation}\label{eq:inted2}
d =\int d\Delta d\omega_a d\omega_b \exp\left(\frac{N^2}{2}\frac{\Delta^2 \Delta_3}{\omega_a\omega_b} +\omega_a(J_a+Q_3) +\omega_b(J_b+Q_3) +2\Delta(Q-Q_3) +2\pi i Q_3\right),
\end{equation}
which is allowed as long as we restrict to the saddle point approximation of the integrand.

The computation of \eqref{eq:inted2} is simplified further since the integrand in the strict EVH limit does not depend on $\omega_b$.
This is due to the fact that the EVH black holes preserve an additional supercharge.
To understand this in more detail, let us first write down the index \eqref{eq:gen-part-funct-choi} after having solved the constraint \eqref{eq:hypersurface-chem-pots}:
\begin{equation}\label{eq:part-funct-choi}
Z(\Delta_I,\omega_i)=\text{Tr}_{\mathcal{H}_{\mathcal{Q}}}\left[ e^{-\Delta_1(Q_1-Q_3)}e^{-\Delta_2(Q_2-Q_3)}e^{-\omega_a (J_a+Q_3)}e^{-\omega_b (J_b+Q_3)}e^{-2\pi i Q_3}e^{-\beta E_{\rm susy}} \right].
\end{equation}
Since EVH black holes have $J_b=Q_3=0$, we can take $\omega_b \to\infty$ in the index without losing their contributions.
Indeed, in this limit the index localizes onto configurations with $J_b+Q_3=0$, of which the EVH black holes form a subset.
The $\omega_b \to\infty$ limit is known as the Macdonald limit of the superconformal index, which can be defined for generic four-dimensional $\mathcal{N}=2$ SCFTs.
One can explicitly check that the supercharge $\mathcal{Q}^{\prime}$ with charges:
\begin{equation}\label{eq:charges-Q'}
Q_1=Q_2=-Q_3=+\frac{1}{2}, \qquad J_a=-J_b=-\frac{1}{2}
\end{equation}
anticommutes with the remaining operator in the trace.
Therefore, the trace localizes on the Hilbert space of 1/8$^{\rm th}$ BPS states annihilated by both $\mathcal{Q}$ and $\mathcal{Q}^{\prime}$.
One may verify the additional charge constraint in the Macdonald sector by noting that:
\begin{equation}
\lbrace \mathcal{Q}^{\prime},\mathcal{S}^{\prime}\rbrace=E-Q_1-Q_2+Q_3-J_a+J_b.
\end{equation}
The vanishing of this expression, combined with the 1/16$^{\rm th}$ BPS condition \eqref{eq:1/16bps-cond}, indeed leads to $J_b+Q_3=0$.

The Macdonald limit of the index was also considered in \cite{Choi:2018hmj}.
Solving the constraint \eqref{eq:hypersurface-chem-pots} in the limit $\omega_b \to\infty$ for $\Delta_3$, one finds that $\Delta_3 \to \infty$ also.
This leaves $\Delta$ and $\omega_a$ unconstrained.\footnote{This claim will be re-examined below. However, even at this stage one should note that only the \textit{real} parts of the chemical potentials $\Delta_{1,2},\omega_a$ are unconstrained, since any finite real part is insignificant with respect to $\omega_b,\Delta_3 \to\infty$.}
In the limit that $\Delta,\omega_a\ll 1$, \cite{Choi:2018hmj} shows that the partition function becomes equal to the naive $\omega_b,\Delta_3 \to \infty$ limit of \eqref{eq:large-n-gen-partn-function}:
\begin{equation}\label{eq:saddleEVHpf}
\log Z= \frac{N^2}{2} \frac{\Delta^2}{\omega_a}.
\end{equation}
Since the charges $J_a$ and $Q$ are of order $\mathcal{O}(N^2)$, we may in complete analogy with the analysis leading to \eqref{eq:blackholeentropy0} compute the Legendre transform of \eqref{eq:saddleEVHpf} with respect to $\Delta$ and $\omega_a$, \textit{i.e.}, we extremize
\begin{equation}
S= \left(\frac{N^2}{2} \frac{\Delta^2}{\omega_a} + \omega_a J_a +2\Delta Q\right)\left|_{\widehat{\Delta},\widehat{\omega}_a}\right.,
\end{equation}
with respect to $\Delta$ and $\omega_a$, where the hatted potentials denote those values at which the extremum is reached. 
One finds that this leads to $S=0$, as expected for an EVH black hole.
However, this extremization procedure only fixes the ratio of $\widehat{\Delta}$ and $\widehat{\omega}_a$ to be:
\begin{equation}
\frac{\widehat{\Delta}}{\widehat{\omega}_a}=-\frac{2Q}{N^2}=-\frac{a}{1-a},
\end{equation}
where we have used the EVH value of $Q=Q_1=Q_2$ given in eq.\ (\ref{eq:non-BPSdata}).
Even though this is consistent with the chemical potentials of the EVH black hole given in \eqref{eq:valueofcem}, we will show that we can do better by correctly taking into account the constraint on $\Delta$ and $\omega_a$.

To continue, we take a different approach to compute the Macdonald limit.
Instead of taking the chemical potential $\omega_b\to \infty$, we will require instead that apart from $\mathcal{Q}$ also $\mathcal{Q}^{\prime}$ anticommutes with the operator in the trace.
This will guarantee that the trace restricts to the Macdonald sector $\mathcal{H}_{\mathcal{Q,\mathcal{Q}^{\prime}}}$ of the theory. 
In doing this, we derive an additional constraint on the chemical potentials, similar to how \eqref{eq:hypersurface-chem-pots} was derived.
We start from the index in which we have solved \eqref{eq:hypersurface-chem-pots} for $\Delta_3$, as given in \eqref{eq:part-funct-choi}.
Demanding that $\mathcal{Q}^{\prime}$ anticommutes with the operator in the trace and using \eqref{eq:charges-Q'}, we derive the following constraint:
\begin{equation}\label{eq:new-constr}
\Delta_1+\Delta_2-\omega_a=2\pi i n \mod 4\pi i, \qquad n\in \mathbb{Z}.
\end{equation}
Combining this with the original constraint, we note that in this case we also have:
\begin{equation}
\Delta_3-\omega_b=2\pi im \mod 4\pi i, \qquad m\in \mathbb{Z} \qquad \text{such that} \qquad m+n\in 2\mathbb{Z}+1.
\end{equation}
Solving the new constraint for $\omega_a$ while taking $n$ odd and $m$ even, we find:
\begin{equation}\label{eq:macdonald-index}
Z(\Delta_1,\Delta_2)=\text{Tr}_{\mathcal{H}_{\mathcal{Q},\mathcal{Q}^{\prime}}}\left[ e^{-\Delta_1(Q_1+J_a)}e^{-\Delta_2(Q_2+J_a)}e^{2\pi i J_a}e^{-\beta E_{\rm susy}} \right].
\end{equation}
Note that we used the fact that $J_b+Q_3=0$ in $\mathcal{H}_{\mathcal{Q},\mathcal{Q}^{\prime}}$ and consequently the index does not depend on $\omega_b$.

We will now work out the same extremization procedure as above, but taking into account the constraint \eqref{eq:new-constr}.
We implement the constraint by a Lagrange multiplier and thus extremize:
\begin{equation}
S= \left(\frac{N^2}{2} \frac{\Delta^2}{\omega_a} + \omega_a J_a +2\Delta Q+\Lambda(2\Delta-\omega_a-2\pi i)\right)\left |_{\widehat{\Delta},\widehat{\omega}_a,\widehat{\Lambda}}\right. .
\end{equation}
This leads again to $S=0$, as expected.
However, due to the additional constraint, we are able to solve for the chemical potentials exactly.
They turn out to be given by:\footnote{There is another branch of solution, which gives an imaginary entropy, and hence is discarded.}
\begin{equation}
\widehat{\Delta}= \frac{2\pi ia}{1+a}, \quad \widehat{\omega}_a = -\frac{2\pi i (1-a)}{1+a}, \quad \widehat{\Lambda}=0.
\end{equation}
Note that these values reproduce the gravity calculation \eqref{eq:valueofcem2} exactly.
For consistency, we should check that these chemical potentials fall within the regime of validity of the analysis in \cite{Choi:2018hmj} leading to \eqref{eq:saddleEVHpf}.
To compare with their analysis, we first shift $\Delta\to \Delta-i\pi$ such that the constraint \eqref{eq:new-constr} reads (for odd $n$):
\begin{equation}
2\Delta-\omega=0\mod 4\pi i. 
\end{equation}
The values of the shifted chemical potentials that extremize $S$ then read:
\begin{equation}\label{eq:extr-chem-pots}
\widehat{\Delta}= \frac{\pi i(a-1)}{1+a}, \quad \widehat{\omega}_a = -\frac{2\pi i (1-a)}{1+a}, \quad \widehat{\Lambda}=0.
\end{equation}
Since the regime of validity is dictated by $|\Delta|,|\omega_a|\ll 1$, we find that our analysis is only consistent for $a\to 1$, or in other words very fast spinning black holes.\footnote{The fast spinning limit of EVH black holes also appeared in \cite{Berkooz:2012qh,Berkooz:2014uwa}, and was crucial in understanding the appearance of a dual chiral two-dimensional CFT. We will further comment on this below.}

Finally, we note that our chemical potentials are purely imaginary, whereas a positive real part is required for convergence of the index.
We take the point of view that these chemical potentials reach $0^+$, as required for convergence, in the EVH limit, as we will further show when considering the near-EVH limit.

\paragraph{The near-EVH limit:}

We will now study the near-EVH limit of the index.
In this case, $J_a$ and $Q_1=Q_2$ are of order $N^2$.
However, the charges $Q_3$ and $J_b$ are of order $N^2 \epsilon^2$, as reviewed in Section \ref{sec:AdS5BH}.
Thus, in the large-$N$ limit, we cannot \textit{a priori} justify a saddle point approximation for the evaluation of the $\omega_b$ integral in \eqref{eq:inted}. 
So let us first perform saddle point approximations for the $\omega_a$ and $\Delta$ integrals in:
\begin{equation}\label{eq:inted-2}
d =\int d\Delta d\omega_a d\omega_b \exp\left(\frac{N^2}{2}\frac{\Delta^2 \Delta_3}{\omega_a\omega_b} +\omega_a(J_a+Q_3) +\omega_b(J_b+Q_3) +2\Delta(Q-Q_3) +2\pi i Q_3\right).
\end{equation}
Understanding the constraint \eqref{eq:hypersurface-chem-pots} as fixing $\Delta_3$ as a function of the other chemical potentials, the saddle point equations for $\Delta$ and $\omega_a$ become respectively:
\begin{align}\label{eq:saddleequation}
\begin{split}
& \frac{N^2}{2} \frac{\Delta \Delta_3 -\Delta^2}{\omega_a\omega_b} +Q-Q_3=0, \\ 
& \frac{N^2}{2} \left( -\frac{\Delta^2\Delta_3}{\omega_a^2\omega_b} +\frac{\Delta^2}{\omega_a\omega_b}\right) +J_a+Q_3 =0.
\end{split}
\end{align}
We will denote the solutions to these equations by $\widehat{\Delta}$ and $\widehat{\omega}_{a}$.
Using the saddle point equations to eliminate the charges $Q-Q_3$ and $J_a+Q_3$, we find: 
\begin{equation}\label{eq:omegabintegrand}
d= \int d\omega_b \exp\left[\frac{N^2}{2} \frac{\widehat{\Delta}^2}{\omega_b} \left(\frac{2\widehat{\Delta}}{\widehat{\omega}_{a}}-1\right) +2\pi i Q_3\right] e^{\omega_b (J_b+Q_3)}.
\end{equation}
To understand how to proceed, we will need to determine the precise expressions corresponding to $\widehat{\Delta}$ and $\widehat{\omega}_{a}$, in particular their dependence on $\omega_b$.
Let us therefore analyze \eqref{eq:saddleequation} in more detail.
First of all, the analysis will be simplified by reparametrizing the chemical potentials in terms of unconstrained variables $z_I$, $I=1,\ldots,4$ \cite{Choi:2018hmj,Hosseini:2017mds}: 
\begin{equation}\label{eq:reparam-chem-pots}
\Delta_I= \frac{2\pi i z_I}{ 1+z_1+z_2 +z_3+z_4}, \quad \omega_a = \frac{-2\pi i}{1+z_1+z_2+z_3+z_4}, \quad \omega_b = -\frac{2\pi i z_4}{1+z_1+z_2+z_3+z_4}.
\end{equation} 
In our case, we use $z_1=z_2=z$, for which the saddle point equations become: 
\begin{align}\label{eq:seriesequation}
\begin{split}
& \frac{N^2}{2} \frac{z (z_3 -z)}{z_4} =Q_3-Q,  \\
& \frac{N^2}{2}\frac{z^2(z_3+1)}{z_4}  =-J_a-Q_3.
\end{split}
\end{align}
Let us first solve for $z_3$ in terms of $z$ and $z_4$.
We find:
\begin{equation}\label{eq:soln-z3}
z_3=-\frac{(J_a+Q)z_4}{\frac{N^2}{2}z(z+1)}.
\end{equation}
This leads to a third order equation to be satisfied by $z$:
\begin{equation}\label{eq:sp-eqn-z}
\frac{N^2}{2}\frac{z^2(z+1)}{z_4}+(Q_3-Q)(z+1)+J_a+Q=0.
\end{equation}
This equation can be solved exactly, but the solutions are not particularly illuminating.
Instead, we will analyze the equation in the near-EVH limit, meaning we take $b=\lambda \epsilon^2$, and take the limit in which $\epsilon\to 0$ and $N\to \infty$ keeping fixed $N^2\epsilon$.
From the gravity solution, we know that $\omega_b\sim \epsilon^{-1}$ for $\epsilon \to 0$, as can be seen from \eqref{eq:BPSomegab}.
In this limit, it turns out that $z_4$ is proportional to $\omega_b$,\footnote{To see this, we note that to the order in epsilon we are considering, $z_3=-z_4+O(z_4^0)$ (as in \eqref{eq:z3in-terms-ofz4}). 
This implies that in the denominator of $\omega_b$ in terms of the $z_I$ \eqref{eq:reparam-chem-pots}, the diverging pieces in $z_3$ and $z_4$ cancel. Therefore, $\omega_b\to \infty$ implies $z_4\to \infty$.} so we may solve \eqref{eq:sp-eqn-z} near $z_4\to \infty$ and keep only finite terms in the limit.
Note that for consistency, we should check that the resulting values of the chemical potentials indeed provide a \textit{dominant} saddle for the integral \eqref{eq:inted-2}.

In the limit $z_4\to \infty$, two of the three solutions to \eqref{eq:sp-eqn-z} blow up, whereas the finite solution reads:
\begin{equation}
z=\frac{a}{1-a}+\frac{a}{(1-a)^2}\frac{1}{z_4}+\mathcal{O}(z_4^{-2}).
\end{equation}
Feeding this back into \eqref{eq:soln-z3}, we also find the solution for $z_3$:
\begin{equation}\label{eq:z3in-terms-ofz4}
z_3=-z_4+\frac{1+a}{1-a}+\mathcal{O}(z_4^{-1}).
\end{equation}
Using \eqref{eq:reparam-chem-pots} and taking $z_4=\frac{i\sqrt{a(1+a)}}{\sqrt{\lambda}\epsilon(1-a)}$, one can verify that these values give the expected chemical potentials for $(\Delta,\omega_a,\omega_b)$ in the near-EVH limit, as featured in \eqref{eq:valueofcem} and \eqref{eq:BPSomegab}:
\begin{align}
\begin{split}
\Delta = \frac{i\pi a}{1+a}+ \mathcal{O}(z^{-1}_4),\quad \omega_a = -\frac{\pi i (1-a)}{1+a}+ \mathcal{O}(z^{-1}_4), \quad \omega_b = \frac{\pi}{\sqrt{\lambda}\epsilon} \sqrt{\frac{a}{1+a}}+ \mathcal{O}(z^{0}_4).
\end{split}
\end{align}

We now plug the solutions into the integrand and take $b=\lambda \epsilon^2$ to obtain:
\begin{equation}\label{eq:omegab-int}
d=\int d\omega_b \exp\left[\frac{N^2}{2}\frac{\pi^2a^2}{1-a^2}\frac{1}{\omega_b}+\frac{N^2\epsilon^2}{2}\frac{\lambda a}{1-a}\omega_b +\mathcal{O}(\epsilon)\right].
\end{equation}
Let us first note that it may seem as if the second term is $\mathcal{O}(\epsilon)$, and we should drop it.
However, since we expect the saddle to have $\omega_b\sim\epsilon^{-1}$, this term is finite in the near-EVH limit and in fact of the same order as the first term.
To justify a saddle point analysis of this integral, large-$N$ is not enough.
This leaves us with two choices: either we fix $N^2\epsilon$ to be a large number, which is allowed in the near-EVH limit, or we take the fast-spinning limit $a\to 1$.
Whichever we choose, we can then solve for the saddle point to find:
\begin{equation}\label{eq:saddle-omegab-int}
\widehat{\omega}_b=\frac{\pi}{\sqrt{\lambda}\epsilon}\sqrt{\frac{a}{1+a}},
\end{equation}
which confirms that the saddle lies at the near-EVH value of $\omega_b$ and in particular shows that the integral for either $N^2\epsilon$ large or $a\to 1$ is dominated by a saddle at large $\omega_b$.
Evaluating the integrand at the saddle then finally gives us the entropy:
\begin{equation}\label{eq:near-evh-entropy}
S=\frac{\pi  a}{1-a}\sqrt{\frac{\lambda a}{1+a}}N^2\epsilon.
\end{equation}
This reproduces exactly the near-EVH entropy computed from gravity \eqref{eq:sbtz}.
Finally, one can check that (\ref{eq:near-evh-entropy}) reproduces, at this order, the near-EVH limit of the  $1/16^{\textrm{th}}$ BPS black hole entropy in (\ref{eq:blackholeentropy0}), with $\epsilon$-expansion of charges
\begin{equation}
\begin{aligned}
&Q_1=Q_2=\frac{a}{2(1-a)}N^2+\frac{ (1+a)}{2(1-a)}\lambda N^2\epsilon^2+\mathcal{O}(\epsilon^4)\,, \qquad  Q_3=-\frac{ a^2}{2(1-a^2)}\lambda N^2 \epsilon^2+\mathcal{O}(\epsilon^4)\,,\\
&J_a=-\frac{a^2}{2(1-a)^2}N^2+\frac{ a(1+3 a +a^2)}{2(1-a)^2(1+a)}\, \lambda N^2\epsilon^2 +\mathcal{O}(\epsilon^4)\,, \quad J_b=\frac{ a (1+2a)}{2(1-a^2)}\, \lambda N^2 \epsilon^2+\mathcal{O}(\epsilon^4)\,.
\end{aligned}
\end{equation}

\paragraph{Connection to Cardy formula:}

As discussed in \cite{Johnstone:2013eg}, the near-EVH entropy can be understood as the entropy of a BTZ black hole that arises in the near-horizon limit of the near-EVH black hole. 
This automatically means the entropy can be written as a Cardy formula~\cite{Cardy:1986ie} for the putative two-dimensional CFT dual.
The central charge and the conformal dimension for the (BPS) case at hand read \cite{Johnstone:2013eg}:
\begin{equation}\label{eq:evh-c-L0}
c=3\sqrt{2}\frac{a^2}{1-a^2}N^2\epsilon, \qquad L_0-\frac{c}{24}=\frac{1}{2\sqrt{2}}\frac{a}{1-a}\lambda N^2\epsilon.
\end{equation}
Before making the connection between the derivation of the Cardy formula and our computation of \eqref{eq:omegab-int}, let us briefly review the derivation of the Cardy formula in two-dimensional CFTs.
Using the fact that the partition function obeys the property $Z(\beta)=Z(\frac{4\pi^2}{\beta})$ and its low temperature expansion reads $Z\approx e^{\frac{\beta c}{24}}$, one has for the high temperature expansion:
\begin{equation}\label{eq:high-T-exp}
Z(\beta)\approx \exp\left(\frac{\pi^2c}{6\beta}\right), \qquad \beta\ll 1.
\end{equation}
The degeneracy at high temperature can then be estimated through a saddle point approximation of the inverse Laplace transform:
\begin{equation}\label{eq:cardy-form-int}
d\approx \int d\beta \exp\left[\frac{\pi^2c}{6\beta}+\beta\left(L_0-\frac{c}{24}\right)\right].
\end{equation}
The saddle lies at:
\begin{equation}\label{eq:saddle-cardy}
\widehat{\beta}=\pi \sqrt{\frac{c}{6\left(L_0-\frac{c}{24}\right)}}=\frac{\sqrt{2}\pi}{\sqrt{\lambda}}\sqrt{\frac{a}{1+a}} ,
\end{equation}
where in the last step we have inserted \eqref{eq:evh-c-L0}.
The saddle point evaluation of \eqref{eq:cardy-form-int} then yields the entropy at high temperature $S=\log d$, and in our case reproduces \eqref{eq:near-evh-entropy}.

The naive regime of validity is as usual $L_0\gg c$, which in this case would correspond to $\lambda \gg 1$, such that the saddle indeed lies at high temperature.
However, it has been established that for ``holographic CFTs'' with $c\gg 1$, the Cardy formula applies already for $L_0\sim c$ (see \textit{e.g.}, \cite{Hartman:2014oaa}, also for a more precise formulation of what defines a holographic CFT).
This regime can be reached by taking either $a\to 1$ or $N^2\epsilon$ large, as can be seen from \eqref{eq:evh-c-L0}.\footnote{
These limits were already encountered when we had to justify the saddle point approximation of the $\omega_b$ integral around \eqref{eq:saddle-omegab-int}.
We will see below that this justification precisely brings us into the extended regime of validity of the Cardy formula for holographic CFTs.}
In particular, this implies that the Cardy formula applies for states with temperature $\widehat{\beta}= \mathcal{O}(1)$ and that the high-temperature expansion \eqref{eq:high-T-exp} extends to $\beta=\mathcal{O}(1)$ up to $\mathcal{O}(c^0)$ corrections \cite{Hartman:2014oaa}.\footnote{
Already in the famous Strominger--Vafa computation \cite{Strominger:1996sh}, the Cardy formula correctly reproduced the entropy when temperatures are strictly speaking ${\cal O}(1)$.
Indeed, the Cardy formula is broadly applicable when there is a supergravity description of the system.
See also \cite{Jejjala:2009if,Belin:2016yll,deLange:2018mri}.}

We are now ready to compare our derivation of the near-EVH entropy from the index with the usual Cardy formula.
For this, let us compare the integrands of \eqref{eq:omegab-int} and \eqref{eq:cardy-form-int} and their respective saddles \eqref{eq:saddle-omegab-int} and \eqref{eq:saddle-cardy}.
We note that if we make the following redefinition:
\begin{equation}\label{eq:omegab-redef}
\omega_b= \frac{\widetilde{\omega}_b}{\sqrt{2}\epsilon},
\end{equation}
the integrand in \eqref{eq:omegab-int} is precisely of the form of \eqref{eq:cardy-form-int} for the near-EVH values of the central charge and conformal dimension quoted in \eqref{eq:evh-c-L0}.
The conclusion we draw from this is that a careful saddle point analysis of the four-dimensional superconformal index for near-EVH black hole charges becomes equivalent to the derivation of a two-dimensional Cardy formula for a two-dimensional holographic CFT.\footnote{See \cite{Cecotti:2015lab} for similar observations in the context of the Schur index.}

Let us say more about the meaning of the redefinition \eqref{eq:omegab-redef}, especially the factor $\epsilon$.
To understand it, we first assume that the two terms in the integrand in \eqref{eq:omegab-int} can be understood as the central charge and conformal dimension of some two-dimensional CFT, \textit{i.e.}:
\begin{equation}\label{eq:c-L0-sym-prod}
\widetilde{c}=3\frac{a^2}{1-a^2}N^2, \qquad \widetilde{L}_0-\frac{\widetilde{c}}{24}=\frac{1}{2}\frac{a}{1-a}\lambda N^2\epsilon^2.
\end{equation}
In the near-EVH limit, $\widetilde{L}_0-\frac{\widetilde{c}}{24}=0$ meaning that we would be looking at a massless BTZ black hole.
However, from the near-horizon geometry, as studied in Section \ref{ssec:near-EVH-bhs}, we know that in addition there is still a conical defect of order $\frac{1}{\epsilon}$, see (\ref{eq:ang-coord-transf}) and (\ref{eq:mbtz}).
Following \cite{deBoer:2010ac}, this results in a new Virasoro symmetry of the boundary theory with a rescaled central charge and fractionated spectrum:\footnote{If the CFT allows a symmetric product description, the conical defect translates into a twist operator in the CFT which projects onto a twisted sector of order $\frac{1}{\epsilon}$.} 
\begin{equation}\label{eq:long-string-transf}
c=\epsilon\, \widetilde{c}, \qquad L_0-\frac{c}{24}=\frac{1}{\epsilon}\,\left(\widetilde{L}_0-\frac{\widetilde{c}}{24}\right).
\end{equation}
Note that the Cardy formula is invariant under such a transformation; this should be understood as the statement that states appearing in the fractionated spectrum provide the dominant contribution to the entropy.
In particular, this gives us the required factors of $\epsilon$ to compare with (the derivation of) the Cardy formula for the near-EVH CFT.

Hence, apart from the factor $\sqrt{2}$,\footnote{From the gravitational point of view, this factor arises due to the specific relation between the AdS$_5$ and AdS$_3$ time coordinates (see \textit{e.g.}, (4.17) and (5.15) in \cite{Johnstone:2013eg}). It is not clear to us presently how this factor could arise from the CFT$_4$ superconformal index. Perhaps like the famous factor of $\frac34$ in the thermal free energy of ${\cal N}=4$ super-Yang--Mills \cite{Gubser:1996de, Gubser:1998nz}, it arises due to strong coupling effects to which our weak coupling analysis is blind.} the redefinition \eqref{eq:omegab-redef} may be understood as resulting from a ``long string phenomenon'' \eqref{eq:long-string-transf} in the IR two-dimensional CFT.
It would be very interesting to find a four-dimensional interpretation of the transformation \eqref{eq:long-string-transf}.
We will further comment on this in the next section and the discussion.

We will conclude with some further comments on the appearance of a fractionated spectrum encountered above.
Such fractionation is well-known to occur for symmetric product CFTs via the so-called long string phenomenon.
Assuming that the EVH CFT is continuously connected to some symmetric product description, we would like to understand the appearance of the factor $N^2$ in the central charge \eqref{eq:c-L0-sym-prod} of the symmetric product CFT.
Since the conical defect is of order $N^2$ as well, it is implied that the symmetric product would have to be of order $N^2$ at least.
Some of the more well-known situations in which symmetric product CFTs arise from gauge theories is in the low energy limit of $N$ D1 branes (matrix string theory) \cite{Dijkgraaf:1997vv} or in the D1-D5 system \cite{Strominger:1996sh}.
Naively, since we have started from just $N$ D3 branes, our setup seems more in the spirit of matrix string theory. 
However, the resulting symmetric product in that case would only be of order $N$, which is related to the fact that the Weyl group of the $SU(N)$ gauge symmetry is $S_N$.
On the other hand, the D1-D5 system has central charge $c\sim Q_1Q_5$. 
Thus taking $Q_{1}Q_{5}\sim N^2$, it would be possible to generate a central charge of $\mathcal{O}(N^2)$.
However, at the moment it is not clear to us if and how an analogy with the D1-D5 system in our present setup would work.\footnote{Perhaps ideas along the lines of \cite{Balasubramanian:2007bs} could be relevant here as well.}

Another, heuristic explanation for the origin of the factor $N^2$ in the central charge is that it arises from the fact that we consider the SYM theory at its maximally deconfining saddle.
This means that all degrees of the freedom of the $SU(N)$ matrix become free, which should somehow translate into the fact that there are $N^2$ independent copies of the two-dimensional CFT.
The symmetric product should then reflect the indistinguishability of these degrees of freedom.

\subsection{$4$d operators for $2$d EVH CFT}\label{sec:letters}

In this section, we would like to make some comments on the question of which operators in the four-dimensional theory contribute to the entropy of the near-EVH black hole.
We begin by reviewing the result of \cite{Berkooz:2014uwa}, who propose an answer to the same question in the strict EVH limit, and then comment on some important differences in the near-EVH case.

Let us first briefly summarize the basic operator content in the $\mathcal{N}=4$ SYM theory.
It contains six scalars ($X,Y,Z$ and conjugates), four (complex) Weyl fermions and two independent components of the gauge field. 
The classification of the superconformal multiplets is denoted as $[k,q,p]_{(j_1,j_2)}$, as first explored in a beautiful paper \cite{Dolan:2002zh}. 
We list all single letter operators subject to the BPS condition $E=J_a+J_b+Q_1+Q_2+Q_3$ in Table \ref{dsingleletters0}, following the notation of \cite{Harmark:2007px}.
These operators may be combined to compose a generic 1/16$^{\rm th}$ BPS operator.
The 1/16$^{\rm th}$ BPS sector is also called the $SU(1,2|3)$ subsector in \cite{Harmark:2007px}.

\begin{table}
	\centering
	\begin{tabular}{|c|c|c|c|c|c|c|}
		\hline 
		Name in \cite{Harmark:2007px}	& $SO(4)[J_a,J_b]$ & Name in \cite{Kinney:2005ej}  & $Q$ & $Q_3$ & $E_0$ & $E$  \\ 
		\hline 
		$Z$ & $[0,0]$ & $Z$  &$ \frac{1}{2}$ & $0$ & $1$ & $1$  \\ 
		\hline 
		$X$ & $[0,0]$ & $X$ & $ \frac{1}{2}$  & $0$ & $1$ & $1$  \\ 
		\hline 
		$W$ & $[0,0]$ & $Y$ & 0 & $1$ &$1$ & $1$  \\ 
		\hline 
		$\bar{F}_+$ & $[1,1]$ & $F_{++}$  & $0$ & $0$ & $2$ & $2$  \\ 
		\hline
		$\chi_1$ & $[\frac{1}{2},-\frac{1}{2}]$ & $\psi_{0,+,+++}$  & $\frac{1}{2}$ & $\frac{1}{2}$ & $\frac{3}{2}$  & $\frac{3}{2}$   \\ 
		\hline 
		$\chi_2$ & $[-\frac{1}{2},\frac{1}{2}]$ &$\psi_{0,-,+++}$  & $\frac{1}{2}$ & $\frac{1}{2}$ & $\frac{3}{2}$ & $\frac{3}{2}$   \\ 
		\hline 
		$\bar{\chi}_3$	& $[\frac{1}{2},\frac{1}{2}]$  & $\psi_{+,0,-++}$ & $0$ & $\frac{1}{2}$ & $\frac{3}{2}$ & $\frac{3}{2}$    \\ 
		\hline 
		$\bar{\chi}_5$	& $[\frac{1}{2},\frac{1}{2}]$  & $\psi_{+,0,+-+}$ & $0$ & $\frac{1}{2}$ & $\frac{3}{2}$ & $\frac{3}{2}$   \\ 
		\hline 
		$\bar{\chi}_7$	& $[\frac{1}{2},\frac{1}{2}]$  & $\psi_{+,0,++-}$ & $\frac{1}{2}$ & $-\frac{1}{2}$ & $\frac{3}{2}$  & $\frac{3}{2}$  \\ 
		\hline 
		$d_1$	& $[1,0]$ & $\partial_{++}$ & $0$ & $0$ & $1$ & $1$   \\ 
		\hline 
		$d_2$	& $[0,1]$ & $\partial_{+-}$ & $0$ & $0$ & $1$ &$1$    \\ 
		\hline 
	\end{tabular} 
	\caption{The single letter operators making up 1/16$^{\rm th}$ BPS operators in \cite{Kinney:2005ej} correspond to the single letter operators comprising the $SU(1,2|3)$ subsector as defined in \cite{Beisert:2004ry,Harmark:2007px}. They satisfy the $E=E_0$ BPS condition, where $E_0$ is the BPS value $E_0=J_a+J_b+2Q+Q_3$, $J_a=J_1+J_2$, $J_b=J_1-J_2$ and $2Q=Q_1+Q_2$.
Note that we are working with the SYM theory on $S^1\times S^3$, dual to AdS$_5$ in global coordinates. In radial quantization, the states on $S^3$ are in one-to-one correspondence with the operators on $\mathbb{R}^{4}$ made up from the letters in the table.}
	\label{dsingleletters0}
\end{table}

Certain operators preserve more supersymmetry. 
In the strict EVH limit, as discussed in Section \ref{ssec:entropy-near-EVH}, the black hole is 1/8$^{\rm th}$ BPS. 
These states are part of the reduced subsector known as the $SU(1,1|2)$ subsector, composed by the single letter operators $(\chi_1,\bar{\chi}_7,\partial_{++},Z,X)$, which satisfy the additional charge constraint $J_b+Q_3=0$. 
Since EVH black holes have $Q_3=J_b=0$, they in fact comprise only a subsector of the $SU(1,1|2)$ sector. 
In \cite{Berkooz:2014uwa}, it is proposed that the EVH black hole is dual to the fermionic subsector of $SU(1,1|2)$, the so-called $SU(1,1)$ subsector, which contains fermions $\chi_1,\bar{\chi}_7$ and the derivative $\partial_{++}$. 
Partial evidence for this relies on the fact that one can construct states using $\chi_1,\bar{\chi}_7,\partial_{++}$ which satisfy the (BPS) EVH condition:
\begin{equation}\label{eq:evh-bps-cond}
\frac{N^2}{2}J_a=Q_1^2.
\end{equation}
This state can be written schematically as \cite{Berkooz:2014uwa}:
\begin{equation}\label{eq:statefermisurface}
\text{Sym} \left[\prod_{a,b=1}^{N^2} \prod_{j=0}^{\frac{K}{2}-1} \psi_j^a \prod_{m=\frac{K}{2}}^{K-1} \bar{\psi}_m^b \right]\,,
\end{equation}
where $\psi_k \equiv \partial_{++}^k \chi_1$ and $\bar{\psi}_k \equiv \partial_{++}^k \bar{\chi}_7$. 
Furthermore, the symmetrization is with respect to $\psi$ and $\bar{\psi}$.
Using the data in Table \ref{dsingleletters0}, one may verify that the charges of this operator are given by: 
\begin{eqnarray} \label{eq:chargeEVH01}
Q_1& =&Q_2 = \frac{N^2K}{2}, \\
Q_3&=& 0, \\
\frac{J_a+J_b}{2} = j_1& =&  N^2\left(\sum_{j=0}^{\frac{K}{2}-1} \frac{j+1}{2} +  \sum_{m=\frac{K}{2}}^{K-1} \frac{m}{2}\right) = \frac{N^2K^2}{4},  \\  \label{eq:fermidata}
\frac{J_a-J_b}{2} =j_2&=& N^2 \left( \sum_{j=0}^{\frac{K}{2}-1} \frac{j}{2} + \sum_{m=\frac{K}{2}}^{K-1} \frac{m+1}{2}\right) = \frac{N^2 K^2}{4}.
\end{eqnarray}
These charges satisfy the condition \eqref{eq:evh-bps-cond} and $J_b=Q_3=0$, as required for the BPS EVH black hole.
Comparing with \eqref{eq:JaQEVH}, we identify:
\begin{equation}
K= \frac{a}{1-a}.
\end{equation}
In \cite{Berkooz:2014uwa}, the authors propose an explicit description of the EVH CFT$_2$ in the large $K$ or fast rotating ($a\to 1$) limit.
In the free limit of the $\mathcal{N}=4$ SYM theory, the two-dimensional CFT is described by free chiral adjoint fermions, or more precisely the (chiral) WZW models:
\begin{equation}
SU(N)_N\oplus SU(N)_N,
\end{equation}
up to a Gauss law constraint projecting onto gauge invariant states.
Including loop corrections for finite Yang-Mills coupling, the primary effect is that the CFT becomes a gauged WZW model.
This provides a specific realization of the fact that the thermodynamics of the near-horizon BTZ black hole can be derived as the IR limit of the thermodynamics of the AdS$_5$ black hole \cite{Johnstone:2013eg}.

For (BPS) near-EVH black holes instead, we have in general $J_b+Q_3\neq 0$, so that we should now take into account in principle the full $SU(1,2|3)$ subsector.
More precisely, a near-EVH black hole has the charges:
\begin{equation}
J_b=\frac{1}{2}\left(\frac{a}{1-a}+\frac{a^2}{1-a^2}\right)\lambda N^2\epsilon^2,\qquad Q_3=-\frac{1}{2}\frac{a^2}{1-a^2}\lambda N^2\epsilon^2.
\end{equation}
Since we keep fixed $N^2\epsilon$, apparently $J_b=Q_3=0$ in the limit.
However, as we have seen in the previous section, the spectrum of the two-dimensional CFT fractionates, as expressed by the long string transformation \eqref{eq:long-string-transf}.
Since $L_0\sim J_b+Q_3$, we denote with $\widetilde{J}_b+\widetilde{Q}_3$ the charges measured with respect to the fractionated spectrum, \textit{i.e.}:
\begin{equation}
\widetilde{J}_b+\widetilde{Q}_3=\frac{1}{2}\frac{a}{1-a}\lambda N^2\epsilon.
\end{equation}
This is finite in the near-EVH limit.
However, presently it is not clear to us how such a fractionation of the spectrum could be described from the perspective of the four-dimensional CFT.
Note that since these charges do not scale with $N^2$, as opposed to $Q$ and $J_a$, it is suggested that we should effectively attach single-trace operators with non-vanishing values for $J_b,Q_3$ to the EVH operator \eqref{eq:statefermisurface} to find the description of states responsible for the near-EVH entropy.
We leave further exploration of these speculations to future work.

\section{Probing EVH-BPS limit in general AdS$_{d+1}$}\label{sec:generaldimension}

In this section, we explore the EVH-BPS limit of black holes in AdS$_{d+1}$ ($d>2$).
It is already known from \cite{Fareghbal:2008eh} that certain AdS$_4$ black holes have an EVH limit, whose field theory dual is proposed to be generalized BMN limit of CFT$_3$.
As in the case of the AdS$_5$ black hole, an AdS$_3$ geometry emerges in the near-horizon limit. 
Other known AdS black hole solutions include AdS$_6$ black holes \cite{Cvetic:2005zi,Chow:2008ip,Choi:2018fdc,Choi:2019miv} and AdS$_7$ black holes \cite{Cvetic:2005zi,Chow:2007ts}. 
An interesting question is whether the success of probing the EVH-BPS limit of AdS$_5$ black hole from the field theory point of view can be extended to general dimensional AdS black holes.
Two situations will appear as we focus on the EVH-BPS limit: 
\begin{itemize}
	\item The black hole reduces to a naked singularity.
	\item An EVH limit of the black hole appears with entropy-temperature scaling $S\sim T^k$ ($1<k<d-1$). 
\end{itemize}
The scaling relation indicates the emergence of a possible AdS$_{k+2}$/CFT$_{k+1}$ correspondence in the far IR of the UV AdS$_{d+1}$/CFT$_{d}$.
We find that only for the supersymmetric AdS$_5$ black holes, one can probe the EVH-BPS limit by using AdS$_3$/CFT$_2$.
We should note that these problems may originate from the fact that for other dimensions than 5, the most general AdS black hole solutions are not (yet) known.

\subsection{Revisiting AdS$_5$}

We have already shown in Section \ref{ssec:evh-limit-bh} that the AdS$_5$ black holes have a well-defined EVH-BPS limit.
However, we will now propose a simpler method of probing this limit without referring to the (complicated) metric. 
The EVH black hole with a pinching AdS$_3$ in its near-horizon is characterized by the scaling relation $S\sim T$. 
However, the situation is a bit more subtle for BPS black holes, which have vanishing temperature. 
The correct way to understand the EVH-BPS black hole in general dimensions is to note that $S/T^k$ is finite for some $k$ as $T\to 0$. 
Therefore, it is essential to understand how temperature and entropy scale as one take near-horizon and BPS limits. 
This requires knowledge of both the non-extremal black hole solutions and their thermodynamics.  

Let us focus on the entropy and temperature in \eqref{eq:non-BPSdata}. 
Approaching the near-horizon region defines $\epsilon$ by $r=\epsilon \rho$. 
For the black hole to have a vanishing entropy, we may take the ansatz \begin{equation}\label{eq:abepsilon}
a\sim \epsilon^{\alpha}, \qquad b\sim \epsilon^{\beta}, \quad r_+\sim \epsilon.
\end{equation}
Without loss of generality, we may assume $\alpha\le \beta $. 
We are interested in the EVH limit with supersymmetry. 
Then $\alpha$ and $\beta$ are constrained by consistency with the BPS limit  \eqref{eq:BPSq} and \eqref{eq:r0ab}. 
The horizon size $r_0^2 \sim ab$ requires $\alpha+\beta =2$. 
Our assumption $a\ge b$ then implies $\alpha \le 1\le \beta$. 
In addition, the BPS value of $q$ \eqref{eq:qBPS} shows that $q \sim \epsilon^\alpha$. Combining these together, we can see in the limit $\epsilon \to 0$ that:
\begin{eqnarray}
S &\sim& \frac{a^2 r_+^2+q r_+^2}{r_+} \sim q r_+ \sim \epsilon^{\alpha+1},
\\
T&\sim& \frac{r_+^4(1+a^2)-a^2b^2}{q r_+^3 } \sim \epsilon^{1-\alpha}.
\end{eqnarray}
The emergence of a near-horizon CFT$_2$ would require $S\sim T$, which fixes $\alpha=0$. 
This is consistent with the solution we studied in Section \ref{sec:AdS5BH}.

\subsection{AdS$_4$}

Certain types of static AdS$_4$ black hole with $U(1)^4$ gauge fields can have  a well-defined EVH limit \cite{Fareghbal:2008eh}. 
We revisit this problem in the case of a supersymmetric black hole. 
We focus on rotating AdS$_4$ black holes \cite{Cvetic:2005zi}. 
The example we will study is the rotating AdS$_4$ black holes have equal pairs of R-charges $Q_1=Q_3$, $Q_2=Q_4$.  
Let us collect some relevant formulas from \cite{Choi:2018fdc}. 
The energy, charges and angular momentum of the black hole are determined by four free parameters $(\delta_i,a,m)$, $i=1,2$, as follows:
\begin{align}
\begin{split}
E &= \frac{m}{2\Xi^2 G}(\cosh 2\delta_1+\cosh 2\delta_2), \quad J=\frac{ma}{2\Xi^2G}(\cosh 2\delta_1+\cosh 2\delta_2), \\
Q_1&= Q_3 = \frac{m}{4G \Xi} \sinh 2 \delta_1, \qquad Q_2=Q_4 =\frac{m}{4G \Xi} \sinh 2 \delta_2.
\end{split}
\end{align} 
The BPS condition requires:
\begin{equation}
e^{2\delta_1+2\delta_2} =1+\frac{2}{a}.
\end{equation} 
Similar to the AdS$_5$ case, further constraints are required to prevent the naked singularity. 
These read: 
\begin{equation} \label{eq:mdelta12}
m^2 = \frac{\cosh^2 (\delta_1+\delta_2)}{e^{\delta_1+\delta_2} \sinh^3 (\delta_1+\delta_2) \sinh(2\delta_1) \sinh(2\delta_2)}.
\end{equation} 
The entropy in the BPS limit satisfies: 
\begin{align}
\begin{split}
 2(Q_1+Q_2)S &= \frac{\pi}{G} J ,\\
 S^2 +\frac{\pi}{G }S-16\pi^2 Q_1 Q_2 &=0.
\end{split}
\end{align} 
The EVH condition forces $J=Q_1Q_2=0$. Without loss of generality, we can assume 
$Q_2=\delta_2=0$. $J=0$
implies $a=0$. 
We then find there is no solution to the singularity free condition \eqref{eq:mdelta12}. 
This means that all the BPS solutions in the EVH limit become solutions with naked singularities, as opposed to the non-BPS solutions of \cite{Fareghbal:2008eh}.

This does not mean there is no hope of studying EVH limit of BPS AdS$_4$ black holes.
In the above EVH limit, we notice that $\delta_2=0$ forces two of four $U(1)$ charges to vanish.  
However, we should expect at least three non-vanishing charges to support a near-horizon AdS$_3$ geometry \cite{Fareghbal:2008eh}. 
Consequently, we expect that studying the EVH limit requires the knowledge of general rotating AdS$_4$ black hole solutions with unequal $U(1)^4$ charges.  Solutions with more general charges were recently found in \cite{Hristov:2019mqp}. It's interesting to revisit EVH-BPS limit in this recently found solution.

\subsection{AdS$_6$}
Another interesting example is the AdS$_6$ black hole solution \cite{Chow:2008ip}. 
A CFT$_5$ account of the entropy has been studied in \cite{Choi:2018fdc}. 
The field theory dual of AdS$_6$ black hole should be an $\mathcal{N}=1$ five-dimensional SCFT. 
The black hole solution has energy, two angular momenta, and one R-charge:
\begin{align}
\begin{split} 
E &= \frac{2\pi m}{3G \Xi_a\Xi_b} \left[ \frac{1}{\Xi_a}+\frac{1}{\Xi_b} + \frac{q}{2m} \left(1+\frac{\Xi_b}{\Xi_a} + \frac{\Xi_a}{\Xi_b}\right)\right] ,\qquad Q= \frac{\pi \sqrt{q^2+2m q}}{G \Xi_a\Xi_b}, \\
J_a &= \frac{2\pi ma}{3G\Xi_a^2 \Xi_b} \left(1+\frac{
	\Xi_bq}{2m}\right), \qquad J_b = \frac{2\pi mb}{3G\Xi_a \Xi_b^2} \left(1+\frac{
	\Xi_aq}{2m}\right), \\
S &= \frac{2\pi^2[(r_+^2+a^2)(r_+^2+b^2)+q r_+]}{3G \Xi_a \Xi_b},\\ 
T&= \frac{2r_+^2(1+r_+^2)(2r_+^2+a^2+b^2)-(1-r_+^2)(r_+^2+a^2)(r_+^2+b^2) +4q r_+^3 -q^2}{4\pi r_+ [(r_+^2+a^2)(r_+^2+b^2)+q r_+]}.
\end{split}
\end{align}
The BPS condition is very similar to what we have discussed for AdS$_5$ black hole:
\begin{equation}
q= \frac{2m}{(2+a+b)(a+b)}.
\end{equation}
The parameter $m$ is related to horizon size $r_+$ via: 
\begin{equation}
2m = \frac{(r_+^2+a^2)(r_+^2+b^2)}{r_+} + \frac{1}{r_+}[r_+(r_+^2+a^2)+q][r_+(r_+^2+b^2)+q].
\end{equation}
We can solve for $q$ in terms of $r_+$ as follows: 
\begin{equation}
q= (a+b+ab)r_+-r_+^3 -i(1+a+b)(r_+^2-r_0^2),
\end{equation}
where $r_0$ is the same as in \eqref{eq:r0ab}. 
Thus the singularity free condition is given by $r_+=r_0$, similar to the AdS$_5$ case. 
The result is: 
\begin{equation}
q= \frac{(1+a)(1+b)(a+b)r_+}{1+a+b}.
\end{equation}
Let us again make the ansatz \eqref{eq:abepsilon} for $a,b$ in the EVH limit. 
It remains true that $\alpha+\beta=2$, since we still have $r^2_0\sim ab$. 
We then find $(r_+^2+a^2)(r_+^2+b^2) \sim a^2r_+^2 \sim \epsilon^{2\alpha +2} $ and $q r_+ \sim \epsilon^{2+\alpha}$. As a result, $S\sim \epsilon^{2+\alpha}$. 
The scaling of temperature is more complicated. 
Each separate term in the temperature scales as:
\begin{align}
\begin{split} 
 2r_+^2 (1+r_+^2)(2r_+^2+a^2+b^2)& \sim \epsilon^{2\alpha+2}, \\
 (1-r_+^2)(r_+^2+a^2)(r_+^2+b^2) &\sim \epsilon^{2\alpha+2}, \\
 4q r_+^3 &\sim \epsilon^{4+\alpha}, \\
  q^2 &\sim \epsilon^{2\alpha+2}, \\
 r_+ [(r_+^2+a^2)(r_+^2+b^2)+q r_+] &\sim \epsilon^{3+\alpha}.
\end{split}
\end{align}
Naively, the leading term in the $\epsilon\to 0$ limit implies that $T\sim \epsilon^{2\alpha+2-(3+\alpha)} =\epsilon^{\alpha-1}$, but this cannot be correct since an extremal black hole should have a vanishing temperature (recall that $\alpha<1$). 

This can be resolved by noting that in the BPS limit, there is a cancellation between the three terms of order $\epsilon^{2\alpha+2}$. 
To see this we should first notice that $q \sim a r_+$. 
One may verify that all the $ r_+^2a^2$ terms in the numerator of $T$ will cancel. 
The subleading term in the numerator of $T$ comes from $r_+^4$ terms, resulting in $T\sim \epsilon^{4-(3+\alpha)} =\epsilon^{1-\alpha}$. 
Setting $\alpha=0$, as we did in AdS$_5$ case, will result in a potential EVH black hole $S\sim T^2$. 
Since $\alpha\le 1$, $S\sim T$ cannot arise for any allowed $\alpha$. 

In summary, the $T \sim \epsilon$ limit $ (\alpha=0)$ can be achieved by   \begin{equation}
q =(a+\lambda b) r_+, \quad r_+\sim \sqrt{b} \sim \epsilon,
\end{equation}
for some general values of $\lambda$. 
The EVH-BPS limit of the AdS$_6$ black hole should be understood as the extremal vanishing horizon limit while keeping $S/T^2$ fixed, which gives rise to an AdS$_4$ factor in the near-horizon limit of the EVH-BPS AdS$_6$ black hole.

\subsection{AdS$_7$}

AdS$_7$ black holes have three angular momenta $J_{1,2,3}$ and two different R-charges $Q_1,Q_2$. 
There are two types of AdS$_7$ black hole solutions known so far. 
The first type is the solution with equal angular momenta \cite{Cvetic:2005zi}. 
The BPS and EVH condition  result in naked singularities, similar to the AdS$_4$ case. 
Thus we do not attempt to study the EVH-BPS limit here. 
The other type is the solution with equal charges but different angular momenta \cite{Chow:2007ts}. 
This case is rather similar to AdS$_6$ discussed above. 
Let us collect some relevant formulas from \cite{Chow:2007ts}. 
The energy, entropy and temperature are given by:
\begin{align}
\begin{split}  
E=& \frac{\pi^2}{8\Xi_1\Xi_2\Xi_3} \left[\sum_i \frac{2m}{\Xi_i} -m +\frac{5q}{2} +\frac{q}{2} \sum_i \left(\sum_{j\neq i} \frac{2\Xi_j}{\Xi_i} -\Xi_i -\frac{2(1+2a_1a_2a_3)}{\Xi_i} \right) \right], \\
S=& \frac{\pi^3}{4\Xi_1\Xi_2\Xi_3 r_+} [(r_+^2+a_1^2)(r_+^2+a_2^2)(r_+^2+a_3^2)+q (r_+^2-a_1a_2a_3)], \\ 
T=& \frac{(1+r_+^2)r_+^2 \sum_i\prod_{j\neq i} (r_+^2+a_j^2) -\prod_i (r_+^2 +a_i^2) +2q (r_+^4+a_1a_2a_3)-q^2}{2\pi r_+ [(r_+^2+a_1^2)(r_+^2+a_2^2)(r_+^2+a_3^2)+q (r_+^2-a_1a_2a_3)] }. 
\end{split}
\end{align}
Just as in the AdS$_{5,6}$ cases, the supersymmetric solution is only singularity free if the horizon size reads:
\begin{equation}
r_{+}^2 =r_0^2 = \frac{a_1a_2+a_2a_3+a_3a_1-a_1a_2a_3}{1-a_1-a_2-a_3}.
\end{equation}
The corresponding value for $q$ is given by:
\begin{equation}
q=- \frac{(1-a_1)(1-a_2)(1-a_3) (a_1+a_2)(a_2+a_3)(a_3+a_1)}{(1-a_1-a_2-a_3)^2}.
\end{equation}
We now make the ansatz: $a_i\sim  \epsilon^{n_i}$ and $n_1\le n_2 \le n_3$. 
The dominant term of $r_0^2$ in the $\epsilon \to 0$ limit is $r_0^2\sim a_1a_2 \sim\epsilon^{n_1+n_2}$. 
So we can derive: 
\begin{equation}
n_1+n_2=2, \qquad n_1\le 1\le n_2\le n_3.
\end{equation}
This implies $q$ should scale as $q \sim a_1^2a_2 \sim \epsilon^{2+n_1}$.
One finds that the entropy in the limit is dominated by:
\begin{equation}
S \sim \frac{1}{r_+} (a_1^2r_+^4 +q r_+^2) \sim q r_+ \sim \epsilon^{3+n_1}.
\end{equation}
Similar to the AdS$_6$ case, each term in the numerator of the temperature can be shown to scale as:
\begin{align}
\begin{split}
 (1+r_+^2)r_+^2 \sum_i\prod_{j\neq i} (r_+^2+a_j^2)  \sim a_1^2 r_+^4 &\sim \epsilon^{2n_1+4}, \\
\prod_i (r_+^2 +a_i^2) \sim   q^2  &\sim \epsilon^{2n_1+4}, \\
 2q (r_+^4+a_1a_2a_3) &\sim \epsilon^{n_1+6} \quad \text{or} \quad \epsilon^{4+n_1+n_3}.
\end{split}
\end{align} 
The dependence on $\epsilon$ of the last term depends on whether $n_3\ge 2$ or $n_3 <2$. 
Again, there is a cancellation between the order $\epsilon^{2n_1+4}$ terms. 
The subleading terms in the numerator come from $r_+^6$, which results in $T\sim \epsilon^{1-n_1}$. 
Setting $n_1=0$ would result in $S\sim T^3$. 
Therefore, a straightforward generalization of EVH-BPS limit in AdS$_7$ black hole cannot give rise to AdS$_3$ black hole in the near-horizon.
However, the result shows that there may well be an AdS$_5$ geometry in the near-horizon limit of the EVH-BPS AdS$_7$ black hole. 

We should note that our result does not imply that AdS$_7$ black hole cannot have an EVH limit with AdS$_3$ appearing in the near-horizon limit. 
Indeed, to fully establish the (non-)existence of an AdS$_3$ factor for the EVH black hole, one needs to know the general AdS$_7$ solutions with unequal charges.

\section{Discussion} \label{sec:discussion}

In this paper, we have employed the recent results for a microscopic enumeration of the entropy of 1/16$^{\rm th}$ BPS black holes in AdS$_5$ from the $\mathcal{N}=4$ $SU(N)$ super Yang-Mills theory \cite{Choi:2018hmj} to study the (near-)EVH/CFT$_2$ correspondence \cite{Johnstone:2013eg}.
Our main result is that the entropy calculation for large AdS$_5$ black holes can be extended to (near-)EVH black holes in the fast rotating limit $a\to 1$ or for $N^2\epsilon$ large.
Moreover, the computation of the Legendre transform of the superconformal index, yielding the black hole entropy, becomes equivalent to the derivation of the Cardy formula in two-dimensional CFT with a central charge and conformal dimension that precisely match the values expected from general considerations of the EVH/CFT$_2$ proposal.
In addition, we have introduced a method to determine the existence of possible BPS EVH black holes in AdS$_{4,6,7}$ and their respective near-horizon geometries.
We found indications that in the cases of AdS$_{6,7}$, the near-horizon geometries contain AdS$_{4,5}$ factors.

There are many directions for further exploration.
First of all, we would like to better understand the underlying mechanism for the appearance of infinite dimensional symmetry algebras, associated with the two-dimensional CFT, from a four-dimensional CFT in the near-EVH limit. 
In particular, the EVH/CFT correspondence implies that a subset of the free operators in four-dimensional $\mathcal{N}=4$ SYM can organize themselves into representations of the Virasoro algebra.
Our derivation of the two-dimensional Cardy formula from the superconformal index sheds some light on this question.
In particular, since the Virasoro zero mode is proportional to $J_b+Q_3$, the operators that account for the near-EVH entropy should lie in the full $SU(1,2|3)$ subsector, as opposed to operators of $SU(1,1|2)$ subsector which comprise the strict EVH black holes. 
Perhaps there exists a connection to \cite{Beem:2013sza}, which derives a chiral algebra from four-dimensional operators through a cohomological construction, and produces the full Virasoro algebra in the BPS near-EVH case we have been studying.
However, this cannot be the complete story because the full $SU(1,2|3)$ subsector is much larger than the Schur $SU(1,1|2)$ sector, which is apparently the only subsector for which the construction of \cite{Beem:2013sza} holds. 
Moreover, the non-unitarity of their chiral algebras is difficult to reconcile with the expected unitarity of the EVH CFT.

In addition, an important element in our derivation of the entropy is the fractionation of the spectrum of the two-dimensional CFT, as implied by the presence of a conical defect.
In two dimensions, such fractionation can be usefully understood in terms of a long string phenomenon \cite{Maldacena:1996ds, Dijkgraaf:1996xw}.
In the four-dimensional CFT, however, a mechanism behind the fractionation is less clear.
It would be very interesting to find an interpretation of the fractionation in four dimensions, which we expect to be possible to achieve at least in the BPS case we have been exploring.

In our investigation of the AdS$_2$ or AdS$_3$ corresponding to EVH black holes, coordinates for the lower dimensional AdS space are realized within the asymptotic AdS$_5$.
In a generic setting, in flowing from AdS$_5\times S^5$ asymptopia to the AdS$_d\times X$ geometry in the near-horizon region, the coordinates from the AdS and sphere mix as we move away from the boundary toward the interior.
(This happens, for example, in \cite{Balasubramanian:2007bs}.)
In other words, generators of the Virasoro algebra in the IR arise as non-trivial combinations of generators of the $SO(2,4)$ conformal symmetry and generators of the $SU(4)$ R-symmetry of the ${\cal N}=4$ super-Yang--Mills theory.
Extending the analysis to such cases is work in progress.

In connecting the UV CFT associated to the asymptotic AdS geometry to the IR CFT associated to the near-horizon geometry, we expect certain $c$-functions are extremized along the RG flow.
This can be made explicit in various contexts \cite{Bhattacharyya:2014gsa} using the attractor mechanism in ${\cal N}=2$ supergravity  \cite{attr1,attr2,attr3,attr4,attr5,attr6,attr7,attr8}.
Perhaps these $c$-functions can help characterize the relation between the Virasoro generators in the IR and the gauge theory operators in the UV.
Indeed, as holographic RG integrates out gravitational degrees of freedom, the $c$-function is a monotonically decreasing function.
However, the entropy of the black hole remains the same from either the UV or the IR perspective; the two are different descriptions of the same system.
Our understanding of entropy is ultimately so far couched in terms of gravitational thermodynamics.
A precise map between microstates in the UV and the IR is necessary to advance the discussion to the statistical physics of gravitational systems.

Finally, the scaling relations $S\sim T^2$ and $S\sim T^3$ in AdS$_{6,7}$ black hole solutions indicate the appearance of $d>2$ dimensional CFTs in their respective EVH limits. 
To understand the physical consequences, we recall what happens if an AdS$_3$ appears in the EVH limit, where $S\sim T$. 
The emergent IR geometry (known as the pinching AdS$_3$) is a locally AdS$_3$ geometry with a pinching angular direction.
In the field theory, such pinching translates into an infinitely gapped system unless the central charge is scaled to infinity \cite{deBoer:2010ac} such that non-trivial dynamics may remain.
For symmetric product CFTs, these dynamics can be traced to the very low energy modes which arise in this limit due to the long string phenomenon, or more generally momentum fractionation \cite{deBoer:2010ac}. 
It is less clear how to think of pinching AdS$_{4,5}$ geometries which arise in (near-)EVH limits of AdS$_{6,7}$ black holes respectively, and whether their dual IR CFT$_{3,4}$ spectra in an appropriate $c\to\infty$ limit exhibit a similar fractionation.
We hope to answer these questions in future work.

\section*{Acknowledgements}
We are grateful for conversations with Pallab Basu, Davide Cassani, Robert de Mello Koch, Seok Kim, James Lucietti, Samir Mathur, and Balt van Rees.
VJ thanks the string group at the University of Pennsylvania for hospitality during his sabbatical when much of this work was performed.
VJ and YL are supported by the South African Research Chairs Initiative of the Department of Science and Technology and the National Research Foundation.
YL is also supported in part by the project “Towards a deeper understanding of black holes with non-relativistic holography” of the Independent Research Fund Denmark (grant number DFF-6108-00340). 
YL also thanks Institute of Theoretical Physics, Chinese Academy of Science of hospitality during various stages of this project.
SvL is supported by the Simons Foundation Mathematical and Physical Sciences Targeted Grants to Institutes, Award ID:509116.
WL is grateful for support from NSFC 11875064, the Max-Planck Partergruppen fund, and the Simons Foundation through the Simons Foundation Emmy Noether Fellows Program at Perimeter Institute.

\appendix


\section{EVH limit of general supersymmetric black holes}\label{sec:generalEVHBH}

When the supersymmetry condition is satisfied, there exists a more general class of black hole solutions of the action \eqref{eq:five-dimensionalaction} with four independent parameters. 
The BPS solution \eqref{eq:nonextremalBH} should be viewed as a two-parameter solution with free parameters ($a,b$). 
The more general solution is given by:
\begin{equation}\label{eq:susyBHansatz}
ds_5^2 = -f^2 (dt+\omega_{\widetilde{\phi}} d\widetilde{\phi} +\omega_{\widetilde{\psi}} d\widetilde{\psi})^2 +f^{-1} h_{mn} dx^m dx^n,
\end{equation}
where $h_{mn}$ is the metric on the base space. 
Supersymmetry requires the base space to be K\"ahler. 
The claim in \cite{Kunduri:2006ek} is that the $h_{mn}$ for solutions in \cite{Chong:2005da} at BPS point is given by:
\begin{align}\label{eq:metricbasespace}
\begin{split}
h_{mn}dx^mdx^n &= (r^2 -r_0^2) \left[\frac{dr^2}{X} + \frac{d\theta^2}{\Delta_\theta} +\frac{\cos^2\theta}{\Xi_b^2}[\Xi_b+\cos^2\theta (\rho^2 +2(1+b)(a+b))] d\widetilde{\psi}^2\right. \\ 
&+ \frac{\sin^2 \theta}{\Xi^2_a}[\Xi_a +\sin^2\theta (\rho^2 +2(1+a)(a+b))] d\widetilde{\phi}^2 \\
&+ \left.\frac{2\sin^2 \theta \cos^2\theta}{\Xi_a \Xi_b} [\rho^2+2(a+b)+(a+b)^2] d\widetilde{\psi}d\widetilde{\phi} \right]\,.
\end{split}
\end{align}
Note that $X(r)$ in the denominator of the $dr^2$ component is exactly as in \eqref{eq:Xr}. 
Therefore, we can compare the metric \eqref{eq:nonextremalBH} to ansatz above. 
One subtlety is that the metric \eqref{eq:susyBHansatz} is in the frame corotating with the horizon. 
It is related to the metric \eqref{eq:nonextremalBH} by the coordinate transformation \eqref{eq:corotatingco}. 
Another subtlety is that the $r_0^2$ dependence of $X(r)$ is redundant, and it can be conveniently set to zero \cite{Kunduri:2006ek}. 
The trick is to realize that one may redefine the parameters $a,b$. 
Let us denote $(a,b,r_0)$ as the set of parameters which after redefinition turn into $(\ba,\bb,0)$. 
It has been shown in \cite{Kunduri:2006ek} that such redefinitions are subject to keeping $A^2,B^2$ fixed, with $A<B$ defined as: 
\begin{equation}
A^2 = \frac{\Xi_a}{\alpha^2}, \quad B^2 =\frac{\Xi_b}{\alpha^2}, \quad \alpha^2 =r_0^2 +(1+a+b)^2.
\end{equation}
For $U(1)^3$ black hole, in new parameters $(\bar{a},\bb,0)$, we have
\begin{align}\label{eq:Himetric}
\begin{split}
H_I &= \frac{\rho^2+3e_I}{r^2} = 1+\frac{\sqrt{\Xi_{\ba}\Xi_{\bb}}(1+ \mu_I) -\Xi_{\ba} \cos^2\theta -\Xi_{\bb}\sin^2\theta}{r^2}, \quad f=(H_1H_2H_3)^{-\frac{1}{3}} ,\\
\omega_\phi &= -\frac{ \sin^2\theta}{ r^2 \Xi_{\ba}} \left[\rho^4 +(2r_m^2+{\ba}^2)\rho^2 +\frac{1}{2}(\beta_2-{\ba}^2{\bb}^2 -({\ba}^2-{\bb}^2))\right] ,\\ 
\omega_\psi &= -\frac{ \cos^2\theta}{ r^2 \Xi_{\bb}} \left[\rho^4 +(2r_m^2+{\bb}^2)\rho^2 +\frac{1}{2}(\beta_2-{\ba}^2{\bb}^2 +({\ba}^2-{\bb}^2))\right] ,
\\ 
r_m^2 &= {\ba}+{\bb} +{\ba}{\bb}, \qquad X^I  = \frac{(H_1H_2H_3)^{\frac{1}{3}}}{H_I}, \quad X(r)= r^2 (r^2+(1+\ba+\bb)^2).
\end{split}
\end{align}
Through a simple calculation, we can show that the BPS limit of solution \eqref{eq:nonextremalBH} can be set in the form of \eqref{eq:susyBHansatz} by an appropriate choice of the $\phi$ and $\psi$ coordinates. 
The conserved charges of the $\frac{1}{16}$-BPS black hole are:\footnote{The $J_\phi$, $J_\psi$ will reduce to $J_a,J_b$ after \eqref{eq:muparameters} conditions are taken. Generically, $a=0$ does not indicate $J_\phi$=0 and similar for $J_b$.} 
\begin{align}\label{eq:supercharges1}
\begin{split}
Q_1 &= \frac{N^2}{2} \left[\mu_1 +\frac{1}{2} (\mu_1\mu_2+\mu_1 \mu_3 -\mu_2\mu_3)\right], \\
Q_2 &= \frac{N^2}{2} \left[\mu_2 +\frac{1}{2} (\mu_3\mu_2+\mu_1 \mu_2 -\mu_1\mu_3)\right],\\
Q_3 &= \frac{N^2}{2}  \left[\mu_3 +\frac{1}{2} (\mu_3\mu_2+\mu_1 \mu_3 -\mu_1\mu_2)\right],\\
\mathcal{J} &\equiv (1+ \mu_1)(1+ \mu_2)(1+ \mu_3) ,\\
J_\phi &= \frac{N^2}{2}  \left[\frac{1}{2}(\mu_1\mu_2+\mu_1 \mu_3 +\mu_2\mu_3) + \mu_1\mu_2\mu_3 +\mathcal{J}  \left(\sqrt{\frac{\Xi_{\bb}}{\Xi_{\ba}}}-1\right)\right], \\ 
J_\psi &= \frac{N^2}{2} \left[\frac{1}{2}(\mu_1\mu_2+\mu_1 \mu_3 +\mu_2\mu_3) + \mu_1\mu_2\mu_3 +\mathcal{J} \left(\sqrt{\frac{\Xi_{\ba}}{\Xi_{\bb}}}-1\right)\right],
\end{split}
\end{align}
under the constraint 
\begin{equation}\label{eq:supersymmetryconstraint}
\mu_1+\mu_2+\mu_3 = \frac{1}{\sqrt{\Xi_{\ba}\Xi_{\bb}}} \left[2r_m^2 +3(1-\sqrt{\Xi_{\ba}\Xi_{\bb}})\right].
\end{equation}
Therefore, this black hole has five parameters $(\mu_I,\ba,\bb)$ with one constraint \eqref{eq:supersymmetryconstraint}.  
The black hole entropy is found to be \cite{Hosseini:2017mds,Kunduri:2006ek,Choi:2018hmj}:
\begin{equation}\label{eq:blackholeentropy}
S = 2\pi \sqrt{Q_1 Q_2 +Q_2 Q_3 +Q_3 Q_1 -\frac{N^2}{2}(J_{\phi}+J_{\psi})}\,.
\end{equation}
The general charge formula can be related to the BPS limit of the equal charge black hole solution \eqref{eq:non-BPSdata} by identifying $\mu_1 =\mu_2 =\mu$, $ \mu_3 =\sigma$ and:\footnote{Note we are using $(a,b,r_0)$ instead of $(\ba,\bb,0)$ for \eqref{eq:non-BPSdata}!} 
\begin{equation}\label{eq:muparameters}
\mu =-1+\frac{(1+a)(1+b)}{\sqrt{\Xi_a\Xi_b}}, \quad \sigma = -1 +\frac{(1+a)(1+b)}{\sqrt{\Xi_a\Xi_b} (1+a+b)}.
\end{equation}
These parameters satisfy:
\begin{equation}
2\mu+\sigma =\frac{1}{\sqrt{\Xi_a\Xi_b}} \Big[2r_m^2+r_0^2+3(1-\sqrt{\Xi_a\Xi_b}) \Big].
\end{equation}
This is consistent with \eqref{eq:supersymmetryconstraint} after reparametrisation. 

The EVH limit of the equal charge black hole \eqref{eq:nonextremalBH} is approached by setting $Q_3=J_\psi=0$, which as shown by \cite{Choi:2018hmj} implies the black hole has zero entropy.  
Then for a more general class of supersymmetric black holes satisfying \eqref{eq:supercharges1}, we can expect to have an AdS$_3$  as the near-horizon geometry after imposing condition $Q_3=J_\psi=0$. 

This turns out to be true. By solving $Q_3=J_\psi=0$, we get:
\begin{align}\label{eq:EVHgeneralSUSY}
\begin{split}
u_1+u_2 &= -2+ \sqrt{\frac{1-\bb^2}{1-\ba^2}} + \frac{3+2\bb+2\ba+2\ba\bb}{\sqrt{(1-\ba^2)(1-\bb^2)}}, \\
u_1 u_2 &=  \frac{(2+2\bb+\bb^2+2\ba+2\ba\bb)(\ba^2-1+\sqrt{(1-\ba^2)(1-\bb^2)})}{(1-\ba^2)^{\frac{3}{2}}(1-\bb^2)^{\frac{1}{2}}}, \\
u_3 &= -1 +\sqrt{\frac{1-\bb^2}{1-\ba^2}}.
\end{split}
\end{align}
To work out the near-horizon geometry, we need to rotate the metric \eqref{eq:susyBHansatz} and \eqref{eq:metricbasespace} to the coordinate system which is static at the horizon. 
Let us define $\hat{\psi} =\widetilde{\psi} - \lambda t $, and look for the value of $\lambda$ which makes the $d\hat{\psi} dt$ term vanish. 
The answer turns our to be $\lambda=1$. 
We verify that under the condition $Q_3=J_\psi=0$, the $d\hat{\psi}d\widetilde{\phi}$ term automatically vanishes. 
Then by this we successfully decouple AdS$_3$ part from AdS$_5$. Then at the end, we are ready to set $r=\epsilon \rho$ and take $\epsilon \to 0$. The final metric of AdS$_5$ in the near-horizon limit is 
\begin{align}
\begin{split}
ds^2 &= h_{\theta} \left( -\frac{\rho^2}{(1-\bb^2)(\ba^2-\bb^2)} \epsilon^2 dt^2 +  \frac{d\rho^2}{\rho^2(1+\ba+\bb)^2} + \frac{\rho^2 }{(1-\bb^2)(\ba^2-\bb^2)} \epsilon^2d\hat{\psi}^2\right), \\
&+ h_\theta \frac{d\theta^2}{ \Delta_\theta} + h_{\theta}f(\theta) d\widetilde{\phi}^2,
\end{split}
\end{align}
where 
\begin{align}
\begin{split}
h_\theta&= \frac{1}{2}\left[(\ba^2-\bb^2)^2\cos^2\theta (8+8\ba+8\bb+8\ba\bb+3\ba^2+\bb^2 \right. \\
&\left. +4(\ba^2+2\ba+2\bb+2\ba\bb+2\bb^2)\cos2\theta + (\ba^2-\bb^2) \cos 4\theta)\right]^{\frac{1}{3}} \\
f(\theta) &= \frac{8(1+\ba+\bb)^2 \Delta_\theta \sin^2\theta}{(1-a^2)^2} \left(8+8\ba+8\bb+8\ba\bb+3\ba^2+\bb^2 \right. \\ 
& \left. +4(\ba^2+2\ba+2\bb+2\ba\bb+2\bb^2)\cos2\theta + (\ba^2-\bb^2) \cos 4\theta \right)^{-1}.
\end{split}
\end{align}
This means that once $Q_3=J_\psi=0$ we may expect an AdS$_3$ geometry in the near-horizon limit even for the most general class of supersymmetric black holes \cite{Kunduri:2006ek}.  

We can expect that the BTZ like geometry can appear as near-EVH limit for the general supersymmetric black hole if we excite the charges $Q_3$ and $J_\psi$. In fact, we can deform the charge constraint \eqref{eq:EVHgeneralSUSY} to be:
\begin{align}\label{eq:EVHgeneralSUSY2}
\begin{split}
u_1+u_2 &= -2+ \sqrt{\frac{1-\bb^2}{1-\ba^2}} + \frac{3+2\bb+2\ba+2\ba\bb}{\sqrt{(1-\ba^2)(1-\bb^2)}} +\lambda \epsilon^2, \\
u_1 u_2 &=  \frac{(2+2\bb+\bb^2+2\ba+2\ba\bb)(\ba^2-1+\sqrt{(1-\ba^2)(1-\bb^2)})}{(1-\ba^2)^{\frac{3}{2}}(1-\bb^2)^{\frac{1}{2}}}, \\
u_3 &= -1 +\sqrt{\frac{1-\bb^2}{1-\ba^2}} -\lambda \epsilon^2.
\end{split}
\end{align}
The new solution also keeps the constraint \eqref{eq:supersymmetryconstraint} unaffected, but makes the charges $Q_3,J_\psi \sim N^2\epsilon^2$. 
We have checked using Mathematica that the near-horizon geometry is BTZ like. 

\bibliographystyle{JHEP}
\bibliography{qec}

\providecommand{\href}[2]{#2}\begingroup\raggedright\begin{thebibliography}{10}

\bibitem{Maldacena:1997re}
J.~M. Maldacena, {\it {The Large N limit of superconformal field theories and
  supergravity}},  {\em Int. J. Theor. Phys.} {\bf 38} (1999) 1113--1133,
  [\href{http://arxiv.org/abs/hep-th/9711200}{{\tt hep-th/9711200}}]. [Adv.
  Theor. Math. Phys.2,231(1998)].

\bibitem{Gubser:1998bc}
S.~S. Gubser, I.~R. Klebanov, and A.~M. Polyakov, {\it {Gauge theory
  correlators from noncritical string theory}},  {\em Phys. Lett.} {\bf B428}
  (1998) 105--114, [\href{http://arxiv.org/abs/hep-th/9802109}{{\tt
  hep-th/9802109}}].

\bibitem{Witten:1998qj}
E.~Witten, {\it {Anti-de Sitter space and holography}},  {\em Adv. Theor. Math.
  Phys.} {\bf 2} (1998) 253--291,
  [\href{http://arxiv.org/abs/hep-th/9802150}{{\tt hep-th/9802150}}].

\bibitem{David:2002wn}
J.~R. David, G.~Mandal, and S.~R. Wadia, {\it {Microscopic formulation of black
  holes in string theory}},  {\em Phys. Rept.} {\bf 369} (2002) 549--686,
  [\href{http://arxiv.org/abs/hep-th/0203048}{{\tt hep-th/0203048}}].

\bibitem{Aharony:2008ug}
O.~Aharony, O.~Bergman, D.~L. Jafferis, and J.~Maldacena, {\it {N=6
  superconformal Chern-Simons-matter theories, M2-branes and their gravity
  duals}},  {\em JHEP} {\bf 10} (2008) 091,
  [\href{http://arxiv.org/abs/0806.1218}{{\tt arXiv:0806.1218}}].

\bibitem{Mori:2014tca}
H.~Mori and S.~Yamaguchi, {\it {M5-branes and Wilson surfaces in
  AdS$_{7}$/CFT$_{6}$ correspondence}},  {\em Phys. Rev.} {\bf D90} (2014),
  no.~2 026005, [\href{http://arxiv.org/abs/1404.0930}{{\tt arXiv:1404.0930}}].

\bibitem{Chester:2018dga}
S.~M. Chester and E.~Perlmutter, {\it {M-Theory Reconstruction from (2,0) CFT
  and the Chiral Algebra Conjecture}},  {\em JHEP} {\bf 08} (2018) 116,
  [\href{http://arxiv.org/abs/1805.00892}{{\tt arXiv:1805.00892}}].

\bibitem{Heckman:2018jxk}
J.~J. Heckman and T.~Rudelius, {\it {Top Down Approach to 6D SCFTs}},  {\em J.
  Phys.} {\bf A52} (2019), no.~9 093001,
  [\href{http://arxiv.org/abs/1805.06467}{{\tt arXiv:1805.06467}}].

\bibitem{Banados:1992wn}
M.~Banados, C.~Teitelboim, and J.~Zanelli, {\it {The Black hole in
  three-dimensional space-time}},  {\em Phys. Rev. Lett.} {\bf 69} (1992)
  1849--1851, [\href{http://arxiv.org/abs/hep-th/9204099}{{\tt
  hep-th/9204099}}].

\bibitem{Banados:1992gq}
M.~Banados, M.~Henneaux, C.~Teitelboim, and J.~Zanelli, {\it {Geometry of the
  (2+1) black hole}},  {\em Phys. Rev.} {\bf D48} (1993) 1506--1525,
  [\href{http://arxiv.org/abs/gr-qc/9302012}{{\tt gr-qc/9302012}}]. [Erratum:
  Phys. Rev.D88,069902(2013)].

\bibitem{Strominger:1997eq}
A.~Strominger, {\it {Black hole entropy from near horizon microstates}},  {\em
  JHEP} {\bf 02} (1998) 009, [\href{http://arxiv.org/abs/hep-th/9712251}{{\tt
  hep-th/9712251}}].

\bibitem{Mathur:2005zp}
S.~D. Mathur, {\it {The Fuzzball proposal for black holes: An Elementary
  review}},  {\em Fortsch. Phys.} {\bf 53} (2005) 793--827,
  [\href{http://arxiv.org/abs/hep-th/0502050}{{\tt hep-th/0502050}}].

\bibitem{Hartman:2014oaa}
T.~Hartman, C.~A. Keller, and B.~Stoica, {\it {Universal Spectrum of 2d
  Conformal Field Theory in the Large c Limit}},  {\em JHEP} {\bf 09} (2014)
  118, [\href{http://arxiv.org/abs/1405.5137}{{\tt arXiv:1405.5137}}].

\bibitem{Kunduri:2007vf}
H.~K. Kunduri, J.~Lucietti, and H.~S. Reall, {\it {Near-horizon symmetries of
  extremal black holes}},  {\em Class. Quant. Grav.} {\bf 24} (2007)
  4169--4190, [\href{http://arxiv.org/abs/0705.4214}{{\tt arXiv:0705.4214}}].

\bibitem{Benini:2015eyy}
F.~Benini, K.~Hristov, and A.~Zaffaroni, {\it {Black hole microstates in
  AdS$_{4}$ from supersymmetric localization}},  {\em JHEP} {\bf 05} (2016)
  054, [\href{http://arxiv.org/abs/1511.04085}{{\tt arXiv:1511.04085}}].

\bibitem{Hosseini:2017mds}
S.~M. Hosseini, K.~Hristov, and A.~Zaffaroni, {\it {An extremization principle
  for the entropy of rotating BPS black holes in AdS$_{5}$}},  {\em JHEP} {\bf
  07} (2017) 106, [\href{http://arxiv.org/abs/1705.05383}{{\tt
  arXiv:1705.05383}}].

\bibitem{Benini:2018ywd}
F.~Benini and P.~Milan, {\it {Black holes in 4d $\mathcal{N}=4$
  Super-Yang-Mills}},  \href{http://arxiv.org/abs/1812.09613}{{\tt
  arXiv:1812.09613}}.

\bibitem{Cabo-Bizet:2018ehj}
A.~Cabo-Bizet, D.~Cassani, D.~Martelli, and S.~Murthy, {\it {Microscopic origin
  of the Bekenstein-Hawking entropy of supersymmetric AdS$_{5}$ black holes}},
  {\em JHEP} {\bf 10} (2019) 062, [\href{http://arxiv.org/abs/1810.11442}{{\tt
  arXiv:1810.11442}}].

\bibitem{Choi:2018hmj}
S.~Choi, J.~Kim, S.~Kim, and J.~Nahmgoong, {\it {Large AdS black holes from
  QFT}},  \href{http://arxiv.org/abs/1810.12067}{{\tt arXiv:1810.12067}}.

\bibitem{Kim:2019yrz}
J.~Kim, S.~Kim, and J.~Song, {\it {A 4d $N=1$ Cardy Formula}},
  \href{http://arxiv.org/abs/1904.03455}{{\tt arXiv:1904.03455}}.

\bibitem{ArabiArdehali:2019tdm}
A.~Arabi~Ardehali, {\it {Cardy-like asymptotics of the 4d $ \mathcal{N}=4 $
  index and AdS$_{5}$ blackholes}},  {\em JHEP} {\bf 06} (2019) 134,
  [\href{http://arxiv.org/abs/1902.06619}{{\tt arXiv:1902.06619}}].

\bibitem{Cabo-Bizet:2019osg}
A.~Cabo-Bizet, D.~Cassani, D.~Martelli, and S.~Murthy, {\it {The asymptotic
  growth of states of the 4d $ \mathcal{N}=1 $ superconformal index}},  {\em
  JHEP} {\bf 08} (2019) 120, [\href{http://arxiv.org/abs/1904.05865}{{\tt
  arXiv:1904.05865}}].

\bibitem{Honda:2019cio}
M.~Honda, {\it {Quantum Black Hole Entropy from 4d Supersymmetric Cardy
  formula}},  {\em Phys. Rev.} {\bf D100} (2019), no.~2 026008,
  [\href{http://arxiv.org/abs/1901.08091}{{\tt arXiv:1901.08091}}].

\bibitem{Lezcano:2019pae}
A.~G. Lezcano and L.~A. Pando~Zayas, {\it {Microstate Counting via Bethe
  Ans\"{a}tze in the 4d ${\cal N}=1$ Superconformal Index}},
  \href{http://arxiv.org/abs/1907.12841}{{\tt arXiv:1907.12841}}.

\bibitem{Larsen:2019oll}
F.~Larsen, J.~Nian, and Y.~Zeng, {\it {AdS$_5$ Black Hole Entropy near the BPS
  Limit}},  \href{http://arxiv.org/abs/1907.02505}{{\tt arXiv:1907.02505}}.

\bibitem{Nian:2019pxj}
J.~Nian and L.~A. Pando~Zayas, {\it {Microscopic Entropy of Rotating
  Electrically Charged AdS$_4$ Black Holes from Field Theory Localization}},
  \href{http://arxiv.org/abs/1909.07943}{{\tt arXiv:1909.07943}}.

\bibitem{Azzurli:2017kxo}
F.~Azzurli, N.~Bobev, P.~M. Crichigno, V.~S. Min, and A.~Zaffaroni, {\it {A
  universal counting of black hole microstates in AdS$_{4}$}},  {\em JHEP} {\bf
  02} (2018) 054, [\href{http://arxiv.org/abs/1707.04257}{{\tt
  arXiv:1707.04257}}].

\bibitem{Bobev:2019zmz}
N.~Bobev and P.~M. Crichigno, {\it {Universal spinning black holes and theories
  of class $ \mathcal{R} $}},  {\em JHEP} {\bf 12} (2019) 054,
  [\href{http://arxiv.org/abs/1909.05873}{{\tt arXiv:1909.05873}}].

\bibitem{Choi:2019zpz}
S.~Choi, C.~Hwang, and S.~Kim, {\it {Quantum vortices, M2-branes and black
  holes}},  \href{http://arxiv.org/abs/1908.02470}{{\tt arXiv:1908.02470}}.

\bibitem{Amariti:2019mgp}
A.~Amariti, I.~Garozzo, and G.~Lo~Monaco, {\it {Entropy function from toric
  geometry}},  \href{http://arxiv.org/abs/1904.10009}{{\tt arXiv:1904.10009}}.

\bibitem{Lanir:2019abx}
A.~Lanir, A.~Nedelin, and O.~Sela, {\it {Black hole entropy function for toric
  theories via Bethe Ansatz}},  \href{http://arxiv.org/abs/1908.01737}{{\tt
  arXiv:1908.01737}}.

\bibitem{Gutowski:2004ez}
J.~B. Gutowski and H.~S. Reall, {\it {Supersymmetric AdS(5) black holes}},
  {\em JHEP} {\bf 02} (2004) 006,
  [\href{http://arxiv.org/abs/hep-th/0401042}{{\tt hep-th/0401042}}].

\bibitem{Gutowski:2004yv}
J.~B. Gutowski and H.~S. Reall, {\it {General supersymmetric AdS(5) black
  holes}},  {\em JHEP} {\bf 04} (2004) 048,
  [\href{http://arxiv.org/abs/hep-th/0401129}{{\tt hep-th/0401129}}].

\bibitem{Chong:2005da}
Z.~W. Chong, M.~Cvetic, H.~Lu, and C.~N. Pope, {\it {Five-dimensional gauged
  supergravity black holes with independent rotation parameters}},  {\em Phys.
  Rev.} {\bf D72} (2005) 041901,
  [\href{http://arxiv.org/abs/hep-th/0505112}{{\tt hep-th/0505112}}].

\bibitem{Chong:2005hr}
Z.~W. Chong, M.~Cvetic, H.~Lu, and C.~N. Pope, {\it {General non-extremal
  rotating black holes in minimal five-dimensional gauged supergravity}},  {\em
  Phys. Rev. Lett.} {\bf 95} (2005) 161301,
  [\href{http://arxiv.org/abs/hep-th/0506029}{{\tt hep-th/0506029}}].

\bibitem{Kunduri:2006ek}
H.~K. Kunduri, J.~Lucietti, and H.~S. Reall, {\it {Supersymmetric multi-charge
  AdS(5) black holes}},  {\em JHEP} {\bf 04} (2006) 036,
  [\href{http://arxiv.org/abs/hep-th/0601156}{{\tt hep-th/0601156}}].

\bibitem{Balasubramanian:2007bs}
V.~Balasubramanian, J.~de~Boer, V.~Jejjala, and J.~Simon, {\it {Entropy of
  near-extremal black holes in AdS(5)}},  {\em JHEP} {\bf 05} (2008) 067,
  [\href{http://arxiv.org/abs/0707.3601}{{\tt arXiv:0707.3601}}].

\bibitem{Fareghbal:2008ar}
R.~Fareghbal, C.~N. Gowdigere, A.~E. Mosaffa, and M.~M. Sheikh-Jabbari, {\it
  {Nearing Extremal Intersecting Giants and New Decoupled Sectors in N = 4
  SYM}},  {\em JHEP} {\bf 08} (2008) 070,
  [\href{http://arxiv.org/abs/0801.4457}{{\tt arXiv:0801.4457}}].

\bibitem{Fareghbal:2008eh}
R.~Fareghbal, C.~N. Gowdigere, A.~E. Mosaffa, and M.~M. Sheikh-Jabbari, {\it
  {Nearing 11d Extremal Intersecting Giants and New Decoupled Sectors in D =
  3,6 SCFT's}},  {\em Phys. Rev.} {\bf D81} (2010) 046005,
  [\href{http://arxiv.org/abs/0805.0203}{{\tt arXiv:0805.0203}}].

\bibitem{SheikhJabbaria:2011gc}
M.~M. Sheikh-Jabbari and H.~Yavartanoo, {\it {EVH Black Holes, AdS3 Throats and
  EVH/CFT Proposal}},  {\em JHEP} {\bf 10} (2011) 013,
  [\href{http://arxiv.org/abs/1107.5705}{{\tt arXiv:1107.5705}}].

\bibitem{deBoer:2011zt}
J.~de~Boer, M.~Johnstone, M.~M. Sheikh-Jabbari, and J.~Simon, {\it {Emergent IR
  Dual 2d CFTs in Charged AdS5 Black Holes}},  {\em Phys. Rev.} {\bf D85}
  (2012) 084039, [\href{http://arxiv.org/abs/1112.4664}{{\tt
  arXiv:1112.4664}}].

\bibitem{Johnstone:2013eg}
M.~Johnstone, M.~M. Sheikh-Jabbari, J.~Simon, and H.~Yavartanoo, {\it
  {Near-Extremal Vanishing Horizon AdS5 Black Holes and Their CFT Duals}},
  {\em JHEP} {\bf 04} (2013) 045, [\href{http://arxiv.org/abs/1301.3387}{{\tt
  arXiv:1301.3387}}].

\bibitem{Sadeghian:2015laa}
S.~Sadeghian, M.~M. Sheikh-Jabbari, M.~H. Vahidinia, and H.~Yavartanoo, {\it
  {Near Horizon Structure of Extremal Vanishing Horizon Black Holes}},  {\em
  Nucl. Phys.} {\bf B900} (2015) 222--243,
  [\href{http://arxiv.org/abs/1504.03607}{{\tt arXiv:1504.03607}}].

\bibitem{deBoer:2010ac}
J.~de~Boer, M.~M. Sheikh-Jabbari, and J.~Simon, {\it {Near Horizon Limits of
  Massless BTZ and Their CFT Duals}},  {\em Class. Quant. Grav.} {\bf 28}
  (2011) 175012, [\href{http://arxiv.org/abs/1011.1897}{{\tt
  arXiv:1011.1897}}].

\bibitem{Berkooz:2014uwa}
M.~Berkooz, P.~Narayan, and A.~Zait, {\it {Chiral 2D "strange metals" from $
  \mathcal{N}=4 $ SYM}},  {\em JHEP} {\bf 01} (2015) 003,
  [\href{http://arxiv.org/abs/1408.3862}{{\tt arXiv:1408.3862}}].

\bibitem{Berkooz:2012qh}
M.~Berkooz, A.~Frishman, and A.~Zait, {\it {Degenerate Rotating Black Holes,
  Chiral CFTs and Fermi Surfaces I - Analytic Results for Quasinormal Modes}},
  {\em JHEP} {\bf 08} (2012) 109, [\href{http://arxiv.org/abs/1206.3735}{{\tt
  arXiv:1206.3735}}].

\bibitem{Beem:2013sza}
C.~Beem, M.~Lemos, P.~Liendo, W.~Peelaers, L.~Rastelli, and B.~C. van Rees,
  {\it {Infinite Chiral Symmetry in Four Dimensions}},  {\em Commun. Math.
  Phys.} {\bf 336} (2015), no.~3 1359--1433,
  [\href{http://arxiv.org/abs/1312.5344}{{\tt arXiv:1312.5344}}].

\bibitem{Basar:2015xda}
G.~Başar, A.~Cherman, K.~R. Dienes, and D.~A. McGady, {\it {4D-2D equivalence
  for large- N Yang-Mills theory}},  {\em Phys. Rev.} {\bf D92} (2015), no.~10
  105029, [\href{http://arxiv.org/abs/1507.08666}{{\tt arXiv:1507.08666}}].

\bibitem{Cvetic:2005zi}
M.~Cvetic, G.~W. Gibbons, H.~Lu, and C.~N. Pope, {\it {Rotating black holes in
  gauged supergravities: Thermodynamics, supersymmetric limits, topological
  solitons and time machines}},
  \href{http://arxiv.org/abs/hep-th/0504080}{{\tt hep-th/0504080}}.

\bibitem{Cvetic:1999xp}
M.~Cvetic, M.~J. Duff, P.~Hoxha, J.~T. Liu, H.~Lu, J.~X. Lu,
  R.~Martinez-Acosta, C.~N. Pope, H.~Sati, and T.~A. Tran, {\it {Embedding AdS
  black holes in ten-dimensions and eleven-dimensions}},  {\em Nucl. Phys.}
  {\bf B558} (1999) 96--126, [\href{http://arxiv.org/abs/hep-th/9903214}{{\tt
  hep-th/9903214}}].

\bibitem{Colgain:2014pha}
E.~Ã. Colgáin, M.~M. Sheikh-Jabbari, J.~F. Vázquez-Poritz, H.~Yavartanoo, and
  Z.~Zhang, {\it {Warped Ricci-flat reductions}},  {\em Phys. Rev.} {\bf D90}
  (2014), no.~4 045013, [\href{http://arxiv.org/abs/1406.6354}{{\tt
  arXiv:1406.6354}}].

\bibitem{Silva:2006xv}
P.~J. Silva, {\it {Thermodynamics at the BPS bound for Black Holes in AdS}},
  {\em JHEP} {\bf 10} (2006) 022,
  [\href{http://arxiv.org/abs/hep-th/0607056}{{\tt hep-th/0607056}}].

\bibitem{Dolan:2008qi}
F.~A. Dolan and H.~Osborn, {\it {Applications of the Superconformal Index for
  Protected Operators and q-Hypergeometric Identities to N=1 Dual Theories}},
  {\em Nucl. Phys.} {\bf B818} (2009) 137--178,
  [\href{http://arxiv.org/abs/0801.4947}{{\tt arXiv:0801.4947}}].

\bibitem{Kinney:2005ej}
J.~Kinney, J.~M. Maldacena, S.~Minwalla, and S.~Raju, {\it {An Index for 4
  dimensional super conformal theories}},  {\em Commun. Math. Phys.} {\bf 275}
  (2007) 209--254, [\href{http://arxiv.org/abs/hep-th/0510251}{{\tt
  hep-th/0510251}}].

\bibitem{Assel:2014paa}
B.~Assel, D.~Cassani, and D.~Martelli, {\it {Localization on Hopf surfaces}},
  {\em JHEP} {\bf 08} (2014) 123, [\href{http://arxiv.org/abs/1405.5144}{{\tt
  arXiv:1405.5144}}].

\bibitem{Assel:2015nca}
B.~Assel, D.~Cassani, L.~Di~Pietro, Z.~Komargodski, J.~Lorenzen, and
  D.~Martelli, {\it {The Casimir Energy in Curved Space and its Supersymmetric
  Counterpart}},  {\em JHEP} {\bf 07} (2015) 043,
  [\href{http://arxiv.org/abs/1503.05537}{{\tt arXiv:1503.05537}}].

\bibitem{Cardy:1986ie}
J.~L. Cardy, {\it {Operator Content of Two-Dimensional Conformally Invariant
  Theories}},  {\em Nucl. Phys.} {\bf B270} (1986) 186--204.

\bibitem{Strominger:1996sh}
A.~Strominger and C.~Vafa, {\it {Microscopic origin of the Bekenstein-Hawking
  entropy}},  {\em Phys. Lett.} {\bf B379} (1996) 99--104,
  [\href{http://arxiv.org/abs/hep-th/9601029}{{\tt hep-th/9601029}}].

\bibitem{Jejjala:2009if}
V.~Jejjala and S.~Nampuri, {\it {Cardy and Kerr}},  {\em JHEP} {\bf 02} (2010)
  088, [\href{http://arxiv.org/abs/0909.1110}{{\tt arXiv:0909.1110}}].

\bibitem{Belin:2016yll}
A.~Belin, J.~de~Boer, J.~Kruthoff, B.~Michel, E.~Shaghoulian, and M.~Shyani,
  {\it {Universality of sparse $d > 2$ conformal field theory at large $N$}},
  {\em JHEP} {\bf 03} (2017) 067, [\href{http://arxiv.org/abs/1610.06186}{{\tt
  arXiv:1610.06186}}].

\bibitem{deLange:2018mri}
P.~De~Lange, A.~Maloney, and E.~Verlinde, {\it {Monstrous Product CFTs in the
  Grand Canonical Ensemble}},  \href{http://arxiv.org/abs/1807.06200}{{\tt
  arXiv:1807.06200}}.

\bibitem{Cecotti:2015lab}
S.~Cecotti, J.~Song, C.~Vafa, and W.~Yan, {\it {Superconformal Index, BPS
  Monodromy and Chiral Algebras}},  {\em JHEP} {\bf 11} (2017) 013,
  [\href{http://arxiv.org/abs/1511.01516}{{\tt arXiv:1511.01516}}].

\bibitem{Gubser:1996de}
S.~S. Gubser, I.~R. Klebanov, and A.~W. Peet, {\it {Entropy and temperature of
  black 3-branes}},  {\em Phys. Rev.} {\bf D54} (1996) 3915--3919,
  [\href{http://arxiv.org/abs/hep-th/9602135}{{\tt hep-th/9602135}}].

\bibitem{Gubser:1998nz}
S.~S. Gubser, I.~R. Klebanov, and A.~A. Tseytlin, {\it {Coupling constant
  dependence in the thermodynamics of N=4 supersymmetric Yang-Mills theory}},
  {\em Nucl. Phys.} {\bf B534} (1998) 202--222,
  [\href{http://arxiv.org/abs/hep-th/9805156}{{\tt hep-th/9805156}}].

\bibitem{Dijkgraaf:1997vv}
R.~Dijkgraaf, E.~P. Verlinde, and H.~L. Verlinde, {\it {Matrix string theory}},
   {\em Nucl. Phys.} {\bf B500} (1997) 43--61,
  [\href{http://arxiv.org/abs/hep-th/9703030}{{\tt hep-th/9703030}}].

\bibitem{Dolan:2002zh}
F.~A. Dolan and H.~Osborn, {\it {On short and semi-short representations for
  four-dimensional superconformal symmetry}},  {\em Annals Phys.} {\bf 307}
  (2003) 41--89, [\href{http://arxiv.org/abs/hep-th/0209056}{{\tt
  hep-th/0209056}}].

\bibitem{Harmark:2007px}
T.~Harmark, K.~R. Kristjansson, and M.~Orselli, {\it {Decoupling limits of N=4
  super Yang-Mills on R x S**3}},  {\em JHEP} {\bf 09} (2007) 115,
  [\href{http://arxiv.org/abs/0707.1621}{{\tt arXiv:0707.1621}}].

\bibitem{Beisert:2004ry}
N.~Beisert, {\it {The Dilatation operator of N=4 super Yang-Mills theory and
  integrability}},  {\em Phys. Rept.} {\bf 405} (2004) 1--202,
  [\href{http://arxiv.org/abs/hep-th/0407277}{{\tt hep-th/0407277}}].

\bibitem{Chow:2008ip}
D.~D.~K. Chow, {\it {Charged rotating black holes in six-dimensional gauged
  supergravity}},  {\em Class. Quant. Grav.} {\bf 27} (2010) 065004,
  [\href{http://arxiv.org/abs/0808.2728}{{\tt arXiv:0808.2728}}].

\bibitem{Choi:2018fdc}
S.~Choi, C.~Hwang, S.~Kim, and J.~Nahmgoong, {\it {Entropy functions of BPS
  black holes in AdS$_4$ and AdS$_6$}},
  \href{http://arxiv.org/abs/1811.02158}{{\tt arXiv:1811.02158}}.

\bibitem{Choi:2019miv}
S.~Choi and S.~Kim, {\it {Large AdS$_6$ black holes from CFT$_5$}},
  \href{http://arxiv.org/abs/1904.01164}{{\tt arXiv:1904.01164}}.

\bibitem{Chow:2007ts}
D.~D.~K. Chow, {\it {Equal charge black holes and seven dimensional gauged
  supergravity}},  {\em Class. Quant. Grav.} {\bf 25} (2008) 175010,
  [\href{http://arxiv.org/abs/0711.1975}{{\tt arXiv:0711.1975}}].

\bibitem{Hristov:2019mqp}
K.~Hristov, S.~Katmadas, and C.~Toldo, {\it {Matter-coupled supersymmetric
  Kerr-Newman-AdS$_4$ black holes}},  {\em Phys. Rev.} {\bf D100} (2019), no.~6
  066016, [\href{http://arxiv.org/abs/1907.05192}{{\tt arXiv:1907.05192}}].

\bibitem{Maldacena:1996ds}
J.~M. Maldacena and L.~Susskind, {\it {D-branes and fat black holes}},  {\em
  Nucl. Phys.} {\bf B475} (1996) 679--690,
  [\href{http://arxiv.org/abs/hep-th/9604042}{{\tt hep-th/9604042}}].

\bibitem{Dijkgraaf:1996xw}
R.~Dijkgraaf, G.~W. Moore, E.~P. Verlinde, and H.~L. Verlinde, {\it {Elliptic
  genera of symmetric products and second quantized strings}},  {\em Commun.
  Math. Phys.} {\bf 185} (1997) 197--209,
  [\href{http://arxiv.org/abs/hep-th/9608096}{{\tt hep-th/9608096}}].

\bibitem{Bhattacharyya:2014gsa}
A.~Bhattacharyya, S.~S. Haque, V.~Jejjala, S.~Nampuri, and A.~V\'eliz-Osorio,
  {\it {Attractive holographic $c$-functions}},  {\em JHEP} {\bf 11} (2014)
  138, [\href{http://arxiv.org/abs/1407.0469}{{\tt arXiv:1407.0469}}].

\bibitem{attr1}
S.~Ferrara, R.~Kallosh, and A.~Strominger, {\it {N=2 extremal black holes}},
  {\em Phys.Rev.} {\bf D52} (1995) 5412--5416,
  [\href{http://arxiv.org/abs/hep-th/9508072}{{\tt hep-th/9508072}}].

\bibitem{attr2}
A.~Strominger, {\it {Macroscopic entropy of N=2 extremal black holes}},  {\em
  Phys.Lett.} {\bf B383} (1996) 39--43,
  [\href{http://arxiv.org/abs/hep-th/9602111}{{\tt hep-th/9602111}}].

\bibitem{attr3}
S.~Ferrara and R.~Kallosh, {\it {Supersymmetry and attractors}},  {\em
  Phys.Rev.} {\bf D54} (1996) 1514--1524,
  [\href{http://arxiv.org/abs/hep-th/9602136}{{\tt hep-th/9602136}}].

\bibitem{attr4}
S.~Ferrara and R.~Kallosh, {\it {Universality of supersymmetric attractors}},
  {\em Phys.Rev.} {\bf D54} (1996) 1525--1534,
  [\href{http://arxiv.org/abs/hep-th/9603090}{{\tt hep-th/9603090}}].

\bibitem{attr5}
A.~Sen, {\it {Black hole entropy function and the attractor mechanism in higher
  derivative gravity}},  {\em JHEP} {\bf 0509} (2005) 038,
  [\href{http://arxiv.org/abs/hep-th/0506177}{{\tt hep-th/0506177}}].

\bibitem{attr6}
D.~Astefanesei, K.~Goldstein, R.~P. Jena, A.~Sen, and S.~P. Trivedi, {\it
  {Rotating attractors}},  {\em JHEP} {\bf 0610} (2006) 058,
  [\href{http://arxiv.org/abs/hep-th/0606244}{{\tt hep-th/0606244}}].

\bibitem{attr7}
A.~Dabholkar, A.~Sen, and S.~P. Trivedi, {\it {Black hole microstates and
  attractor without supersymmetry}},  {\em JHEP} {\bf 0701} (2007) 096,
  [\href{http://arxiv.org/abs/hep-th/0611143}{{\tt hep-th/0611143}}].

\bibitem{attr8}
A.~Sen, {\it {Black Hole Entropy Function, Attractors and Precision Counting of
  Microstates}},  {\em Gen.Rel.Grav.} {\bf 40} (2008) 2249--2431,
  [\href{http://arxiv.org/abs/0708.1270}{{\tt arXiv:0708.1270}}].

\end{thebibliography}\endgroup

\end{document}